\documentclass[pre,10pt,showkeys,showpacs,twocolumn,superscriptaddress]{revtex4-1}
\usepackage[UKenglish]{babel}
\usepackage{amsmath,amssymb}
\usepackage{fullpage}
\usepackage{bm}
\usepackage{bbold}
\usepackage{appendix}
\usepackage{graphicx}
\usepackage{pstricks}
\usepackage{verbatim}
\usepackage{color}
\usepackage{tikz}
\pgfrealjobname{KalmanFilterPRE_fin}
\usetikzlibrary{calc,decorations.markings}
\newcommand{\be}{\begin{equation}}
\newcommand{\ee}{\end{equation}}
\allowdisplaybreaks[1]
\newcommand{\ac}{\mathcal{C}}

\newcommand{\cf}{\bm{C}_{\rm f}}

\newcommand{\cb}{\bm{C}_{\rm b}}
\newcommand{\cbi}{\bm{C}^{-1}_{\rm b}}

\newcommand{\cfsi}{\bm{C}_{\rm f}^{-1}}
\newcommand{\ksb}{\bm{K}^{\rm sb}}
\newcommand{\kbs}{\bm{K}^{\rm bs}}
\newcommand{\kbbs}{\bm{K}^{\rm bb|s}}
\newcommand{\ekbbs}{r}
\newcommand{\B}{D}
\newcommand{\jm}{\bm{K}}
\newcommand{\omr}{\Omega}
\begin{document}
\title{Inferring hidden states in Langevin dynamics on large networks: 
Average case performance}
\author{B. Bravi}
\email{barbara.bravi@kcl.ac.uk}
\affiliation{Department of Mathematics, King's College London, Strand, London, WC2R 2LS UK}
\author{M. Opper}
\affiliation{Department of Artificial Intelligence, Technische Universit\"{a}t Berlin, Marchstra{\ss}e 23,
Berlin 10587, Germany}
\author{P. Sollich}
\affiliation{Department of Mathematics, King's College London, Strand, London, WC2R 2LS UK}
\setcounter{page}{1}
\pagenumbering{arabic} 
\begin{abstract}
We present average performance results for dynamical inference problems in large networks, where a set of nodes
is hidden while the time trajectories of the others are observed. Examples of this scenario can occur in signal transduction and gene regulation networks.
We focus on the linear stochastic dynamics of continuous variables interacting via random Gaussian couplings of generic symmetry. 
We analyze the inference error, given by the variance of the posterior distribution over hidden paths, in the thermodynamic 
limit and as a function of the system parameters and the ratio $\alpha$ between the number of hidden and observed nodes.
By applying Kalman filter recursions we find that the posterior dynamics is governed by 
an ``effective'' drift that incorporates the effect of the observations. 
We present two approaches for characterizing the posterior variance that allow us to tackle, respectively,
equilibrium and non-equilibrium dynamics. The first appeals to Random Matrix Theory and reveals average spectral properties of 
the inference error and typical posterior relaxation times, the second is based on dynamical functionals and yields the inference error as the solution of an algebraic equation.
\end{abstract}

\maketitle

\section{Introduction}

Inferring the time evolution of a partially observed system of continuous degrees of freedom (d.o.f.)
is an important problem in statistical physics. In systems biology these d.o.f.\ might for example be concentrations of interacting molecular species in biochemical networks.
Inference of unobserved or hidden d.o.f.\ is then
often crucial, e.g.\ for an understanding of molecular mechanisms underlying 
genetic and metabolic processes. Hidden d.o.f.\ can occur because the behavior of part of a network is simply not recorded,
or because the amount of experimental data available might be limited \cite{braunstein}.
If as in our analysis one studies generic continuous d.o.f., a potentially broad 
and interdisciplinary range of applications can be envisaged beyond biology, e.g.\ in financial data \cite{tsay} or weather forecasting \cite{meteo}.

Inference has been studied using statistical mechanics approaches predominantly in scenarios without a temporal dimension, e.g.\ when 
learning from examples in neural networks 
\cite{engel,opperdomany}. Several studies have, like ours, focused on performance analysis in the thermodynamic limit of large systems \cite{seung,sollich}.
Especially for linear learning problems, the spectrum of the input correlation matrix (or equivalently the average response function) 
has turned out to be a key quantity and has been studied by different means, including the replica method \cite{solla,opperdomany,oppersolvable} based on the 
pioneering work of \cite{edjon}, diagrammatic techniques \cite{hertz} and partial differential equations from matrix identities \cite{sollich}.
A key system parameter is the ``storage'' ratio between the number of training examples and the number of parameters to be learned \cite{solla,hertz}.

Rather less work has been done for inference based on entire temporal trajectories, with most efforts focused on the dynamics of discrete variables, typically Ising spins with random asymmetric couplings: see \cite{romanobattistin} for a 
review and \cite{inference1,inference2,roudihertz,roudi1} for examples. 
We extend these studies significantly by accounting for generic interaction symmetry, thus allowing us to interpolate across a range of non-equilibrium situations all the way to equilibrium dynamics. The results we present are exact in the thermodynamic limit and complement our previous study using an a priori approximate method, the Extended Plefka Expansion 
\cite{plefkaobs,plefkaobs2}. 
Our emphasis on non-equilibrium dynamics is motivated by the fact that many biological processes are out of equilibrium. Indeed,
recent studies \cite{berg} and computational models \cite{sanguinetti2} have called for a non-equilibrium approach to 
gene expression dynamics that would allow one to infer regulatory interactions and transcription factor activity from time-resolved measurements.

We focus on a paradigmatic scenario: stochastic linear dynamics on a network of continuous d.o.f.\ that interact via random Gaussian couplings. 
Such linear dynamics should give a reasonable account also of the behavior of generic nonlinear networks of continuous d.o.f.\ near stable fixed points.
We show that our setting is closely related to (linear Gaussian) state space modelling in statistics~\cite{bishop}, where the dynamics of a set of hidden variables can only be observed indirectly. 
This allows us to deploy 
inference methods developed for such models \cite{anderson, vinter, bishop}, specifically the Kalman filter (and smoother)~\cite{kalmanor}.

The distribution over network trajectories is Gaussian in our setting, 
and hence so is the posterior over hidden trajectories given a time trajectory of the observed nodes, as we will make clear. Its mean gives the optimal prediction of the time-dependent hidden state, while the second order statistics give information on the certainty of this prediction. In particular, 
the normalized trace of the equal-time posterior covariance matrix will be our measure of inference error. Posterior covariances between different times quantify temporal correlations of prediction uncertainties.

The novelty of our approach is that we assess the inference error of the Kalman filter for \emph{random} interactions, which induce a random distribution in the eigenvalues of
the posterior covariance. In the thermodynamic limit of large networks that we consider, the spectrum becomes self-averaging: its fluctuations tend to zero, and it becomes equal to the disorder (random interaction) average of the spectrum. We tackle this disorder average 
by exploiting Random Matrix Theory (RMT) results \cite{mehta}. For related approaches that connect RMT and Bayesian statistics see
\cite{vakili1,bun} and references therein.

We will see that the combination of Kalman filter and RMT gives a wealth of information for inference in systems with equilibrium dynamics, i.e.\ obeying detailed balance, but cannot be extended in an obvious way to non-equilibrium dynamics. For these scenarios we choose an alternative avenue, using dynamical functionals and defining the normalization factor of the posterior as a partition function. Again we consider the disorder average, for which in our case an annealed approximation is sufficient instead of a replica treatment.
The replica approach was used for inference of spins trajectories in~\cite{inference1} generalizing to dynamics an approach 
that was already used for learning in static networks (see \cite{seung, engel, opperdomany}). 

The aim of this paper is to provide exact results on the average inference error for large
size networks, against which other approximation methods or algorithms, 
can be compared. Exactness in the thermodynamic 
limit relies crucially on the assumption of weak long-range (mean field) interactions. In addition to the use of Kalman filter recursions combined with RMT, as well 
as dynamical functionals, we provide a link to variational methods.

The paper is organized as follows.
After presenting the governing Kalman filter equations for the posterior variance and the effective posterior drift (section \ref{Stationary_Posterior_Variance}), we use RMT to 
study the equilibrium dynamics case in section \ref{sec:RMT}, first for the elementary case of hidden variables with only self-interactions (section \ref{sec:selfinteraction}), then for 
symmetric hidden-hidden couplings (section \ref{sec:symmetric_couplings}), where we apply free probability methods.
Moving on to non-equilibrium dynamics, we describe in section \ref{Dynamical_functional} the dynamical functional method. We focus on the fully asymmetric case
(section \ref{sec:asydyn}) initially, which then generalizes to 
arbitrary symmetry (section \ref{sec:gensym}). The result is an algebraic equation for the stationary posterior variance in the Laplace
domain which coincides with the one we derived using the Extended Plefka Expansion in \cite{plefkaobs,plefkaobs2}. We summarize and discuss the outlook for future 
work in section~\ref{Discussion}.

\section{Model and general expression for posterior covariance}
\label{Stationary_Posterior_Variance}
The setting we study consists of two sets of variables: the
subnetwork, which models the \emph{observed} d.o.f.\, and the bulk,
which stays \emph{hidden} and whose values we want to infer from the observations.
To allow explicit insight into how the level of accuracy in this inference task depends on the structural parameters of the problem we consider a tractable scenario, where
subnetwork and bulk interact \emph{linearly}.

Our model, then, is a linear dynamical system specified by the following equations 
\begin{eqnarray}
\partial_t \bm{x}^{\rm b}(t)&=&\kbs\bm{x}^{\rm s}(t)+\jm^{\rm bb}\bm{x}^{\rm b}(t)+\bm{\xi}^{\rm b}(t)\label{eq:lineq1}\\
\partial_t \bm{x}^{\rm s}(t)&=&\jm^{\rm ss}\bm{x}^{\rm s}(t)+\ksb\bm{x}^{\rm b}(t)+\bm{\xi}^{\rm s}(t),\label{eq:lineq2}
\end{eqnarray}
where subnetwork and bulk variables are denoted respectively by the superscript $\rm s$ and $\rm b$;
$\bm{\xi}^{\rm s}(t)$ and $\bm{\xi}^{\rm b}$ are independent white Gaussian noises with zero mean and variance
\begin{eqnarray}
\langle\bm{\xi}^{\rm s}(t)\bm{\xi}^{\rm s}(t')^{T}\rangle&=&\bm{\Sigma}^{\rm ss}\delta(t-t')
\\
\langle\bm{\xi}^{\rm b}(t)\bm{\xi}^{\rm b}(t')^{T}\rangle&=&\bm{\Sigma}^{\rm bb}\delta(t-t').
\end{eqnarray}
In addition the matrix $\jm^{\rm ss}$ ($\jm^{\rm bb}$) contains the linear couplings between subnetwork (bulk) 
variables while $\kbs, \ksb$ specify the interactions between subnetwork and bulk.

As pointed out in the introduction, a linear system with Gaussian noise produces a Gaussian distribution over the dynamical 
trajectories of the entire network. By this we mean that the collection of trajectories of all variables is a Gaussian process: the joint distribution of
any finite collection of variables $\lbrace x_i(t_j)\rbrace$  is a multivariate Gaussian. To make this more intuitive it can be helpful to think about
a time discretized version of the dynamics  \eqref{eq:lineq1} and \eqref{eq:lineq2}, 
for which the joint distribution of the collection of subnetwork and bulk variables across all time steps is then Gaussian, as also shown in appendix \ref{appendix:kalman}. 
Inferring the hidden dynamics then corresponds to Gaussian conditioning. In particular, 
the aim is to evaluate the posterior probability distribution over  hidden trajectories, conditioned on the observed subnetwork trajectory.
We denote the latter $\bm{X}^{\rm s}$, as a shorthand for the data sequence $\lbrace \bm{x}^{\rm s}(t)| t \in [0,T]\rbrace$.
The posterior distribution is then fully characterized by the first and second moments
\begin{eqnarray}
\langle \bm{x}^{\rm b}(t)\rangle &=& \bm{\mu}^{\rm b}(t)\\
\langle \delta \bm{x}^{\rm b}(t)\delta \bm{x}^{\rm b}(t')^{T}\rangle 
&=& \bm{C}^{\rm bb|s}(t,t'),
\end{eqnarray}
where $\delta \bm{x}^{\rm b}(t)= \bm{x}^{\rm b}(t)- \bm{\mu}^{\rm b|s}(t)$ is the deviation from the posterior mean and
the $T$ superscript denotes vector or matrix transpose. As defined, $\bm{C}^{\rm bb|s}(t,t)$ is then the posterior covariance matrix of $\bm{x}^{\rm b}(t)$.
We shall drop the superscripts for the sake of brevity so will denote $\bm{\mu}^{\rm b|s}(t)$ simply by $\bm{\mu}(t)$
and $\bm{C}^{\rm bb|s}(t,t')$ by $\bm{C}(t,t')$. The 
best estimate -- in the mean-square sense -- of the hidden dynamics based on the observed time 
series $\bm{X}^{\rm s}$ is then just $\bm{\mu}(t)$, while $\bm{C}(t,t)$ determines the uncertainty in this prediction: in particular, the trace of 
$\bm{C}(t,t)$ is the total mean squared prediction error for the hidden variables. Normalizing by the number of hidden nodes defines what we will call 
the inference error.

To find the posterior means and variances in linear-Gaussian state models one can use 
a message passing algorithm known as \emph{Kalman Filter} \cite{kalmanor} (see appendix \ref{appendix:kalman}).
For a long time series, the algorithm will converge to stationary values for the covariances when well away from the two ends $t=0$ and $t=T$;
note though that the state prediction $\bm{\mu}(t)$ remains time dependent as it is driven by the time dependence of the observed 
$\bm{x}^{\rm s}(t)$. The covariances, on the other hand, are entirely independent of the $\bm{x}^{\rm s}(t)$, by a general property of conditional Gaussian distributions: 
they depend only on which variables are observed, but not their values. 
Note that this contrasts with the case of e.g.\ binary spins, where mean and variance are directly related so that variances of individual spins would generally also be non-stationary.

The stationary inference error, i.e.\ the normalized trace of the stationary equal time posterior covariance $\bm{C}(t,t)=\bm{C}$, will be the main focus of our attention. As shown in appendix \ref{appendix:kalman}, $\bm{C}$ satisfies
\begin{equation}
\label{eq:post}
 \jm^{\rm bb|s}\bm{C}+ \bm{C}\jm^{\rm bb|s\,\it{T}}+\bm{\Sigma}^{\rm bb}=0.
\end{equation}
This is a Lyapunov equation with an ``effective" or ``posterior" drift
$\jm^{\rm bb|s}$, where we use the superscript $\rm bb|s$ to indicate that this is the bulk-bulk coupling matrix conditioned on the observed 
subnetwork trajectory. By ``posterior" we mean then that $\jm^{\rm bb|s}$ incorporates the effect of the observations and defines an effective posterior dynamics
\be
\label{eq:efflineq1}
\partial_t \delta \bm{x}^{\rm b}(t)=\jm^{\rm bb|s}\delta\bm{x}^{\rm b}(t)+\bm{\xi}^{\rm b}(t).
\ee 
The effective drift can be written as
\be
\label{eq:postdrift}
\jm^{\rm bb|s}= \jm^{\rm bb} - \bm{\Sigma}^{\rm bb}\bm{A},
\ee
where $\bm{A}=\bm{A}^{T}$ is a symmetric matrix that is a solution of 
the matrix Riccati (i.e.\ quadratic) equation
\begin{equation}
\label{eq:riccati1}
 \bm{A}\bm{\Sigma}^{\rm bb}\bm{A}-\bm{A}\jm^{\rm bb}-\jm^{\rm bb\,\it{T}}\bm{A}=\bm{W}.
\end{equation}
Here the \emph{feedback} matrix $\bm{W}=\bm{K}^{\rm sb\,\it{T}}(\bm{\Sigma}^{\rm ss})^{-1}\ksb$ 
describes how observations affect the inferred statistics.
This matrix is determined by the interplay between the strength of hidden-observed interactions $\ksb$ and the 
dynamical noise on the observed variables, namely $\bm{\Sigma}^{\rm ss}$. 
(We stress here that this is noise acting on the time evolution of $\bm{x}^{\rm s}$, {\em not} noise affecting our measurement of the observed trajectory.)

The matrix $\bm{A}$ in \eqref{eq:postdrift} is directly related to the backwards messages sent in the Kalman filter method. 
Specifically, the distribution of $\delta \bm{x}^{\rm b}(t)$ conditioned only on observations {\em from time $t$ onwards} is Gaussian, and $\bm{A}$ is its inverse 
covariance in the stationary regime.

Accordingly, equation \eqref{eq:riccati1} can
be derived as the stationary limit of what is known as a Riccati recursion, for the backward pass  in the Kalman Filter (see appendix \ref{appendix:kalman}).
Without observations the distribution of $\bm{x}^{\rm b}(t)$ conditional only on data beyond $t$ is flat, hence $\bm{A}$ vanishes. Then $\kbbs$ reduces to $\jm^{\rm bb}$ as expected and the posterior covariance to the unconditional covariance because \eqref{eq:post} becomes simply $\jm^{\rm bb}\bm{C}+ \bm{C}\jm^{\rm bb\,\it{T}}+\bm{\Sigma}^{\rm bb}=0$. 
One sees therefore that $\bm{A}$ is the key quantity that captures the effects of the observations on the (second order) posterior statistics. 
This insight is supported by an alternative variational derivation of \eqref{eq:post}, \eqref{eq:postdrift} and \eqref{eq:riccati1}, outlined in Appendix 
\ref{appendix:variational}, where $\bm{A}$ appears as a Lagrange multiplier implementing the constraints resulting from the observed data. 

Once the stationary equal-time covariance $\bm{C}$ has been found, it is clear from \eqref{eq:efflineq1} that the two-time covariance must be given by
\begin{equation}
\label{postdecay}
\bm{C}(t-t')= e^{\bm{K}^{\rm bb|s}(t-t')}\bm{C}
\end{equation} 
for $t>t'$. This exponential decay with the effective drift matrix $\bm{K}^{\rm bb|s}$ can be derived explicitly by
generalizing the filtering-smoothing procedure (see appendix \ref{appendix:kalman} and references there).
We have emphasized in the notation the fact that $\bm{C}(t-t')$ depends only on the time difference because the stationary regime obeys time-translation invariance. 
Stability of the conditional 
hidden dynamics, where \eqref{postdecay} decays to zero as $t-t'$ grows, requires $\bm{K}^{\rm bb|s}$ to be negative definite. Assuming that the dynamical matrix 
$\bm{K}^{\rm bb}$ of the isolated hidden dynamics has this property, then also $\bm{K}^{\rm bb|s}$ does because $\bm{A}$, as the inverse covariance matrix in the stationary backwards messages, is non-negative definite.

So far in this section we have derived expressions for $\bm{C}$ and $\bm{C}(t-t')$ that specify the second order posterior statistics in our 
setting of inferring hidden state trajectories. These results are valid for given values of the interaction matrices $\bm{K}^{\rm bb}$ etc. In the 
remainder of the paper we consider these interactions to be drawn from some probability distribution, acting as \emph{quenched disorder}. In an 
appropriately defined infinite size or thermodynamic limit we then expect key results such as the eigenvalue spectrum of $\bm{C}$ to be self-averaging, 
i.e.\ independent of the specific realization. In particular we look at a fully connected system interacting via Gaussian couplings. 
This is a standard scenario used to analyze 
the mean-field regime of e.g.\ spin glass models \cite{sompolinsky1}. It can also be thought of as the large connectivity limit
of an Erd\H{o}s-R\'enyi graph \cite{ER} with Gaussian weights \cite{Rogers}; studying dynamical processes on such random graphs 
to predict the evolution of each node from partial observations is of interest in e.g.\ epidemic forecasting \cite{dallasta1,dallasta2}.
A precedent for the use of RMT techniques, such as Stieltjes transforms and free probability, 
in the study of asymptotic eigenvalue distributions for random Lyapunov and Riccati recursions -- like those occuring in filtering -- can be found in \cite{vakili1}. 
Ref. \cite{vakili1} takes a control and systems theory perspective, however, while we focus on inference for dynamics. 
It is worth stressing that this makes our approach more general, as we look at a time dependent problem with quenched, ``frozen'' randomness 
rather than a sequence of signals where the randomness in the interactions is re-sampled at each step.
From the spectrum $\bm{C}$ we will obtain the inference error; we will also study the properties of the posterior drift $\bm{K}^{\rm bb|s}$, 
whose inverse defines the spectrum of relaxation times of the posterior dynamics.

\section{Thermodynamic limit by Random Matrix Theory}
\label{sec:RMT}
To investigate the thermodynamic limit, we first apply tools from random matrix theory (RMT) to {\em equilibrium} dynamics, where detailed balance holds. We study two such scenarios. 
In the first, the hidden variables only have self-interactions (Sec.~\ref{sec:selfinteraction}); in the second we add random symmmetric hidden-to-hidden interactions 
(Sec.~\ref{sec:symmetric_couplings}). The main results are explicit mathematical expressions which establish a link between the inference error 
and the parameters describing the dynamics.
In both cases we make the same assumptions regarding the hidden-to-observed interactions $\bm{K}^{\rm sb}$, and therefore discuss first the resulting statistics of the feedback matrix $\bm{W}$.

\subsection{Feedback matrix: Wishart ensemble}

The feedback matrix $\bm{W}=\bm{K}^{\rm sb\,\it{T}}(\bm{\Sigma}^{\rm ss})^{-1}\ksb$ is a positive definite symmetric matrix of size $N^{\rm b} \times N^{\rm b}$, 
where $N^{\rm b}$ is the number of hidden variables, i.e.\ the number of components of the vector ${\bm x}^{\rm b}$. We assume throughout in the following that the elements of the $N^{\rm s} \times N^{\rm b}$ matrix
$\ksb$ are independent zero mean Gaussian random variables of fixed variance $k^2/N^{\rm b}$. If $\bm{\Sigma}^{\rm ss}=\sigma_{\rm s}^2\mathbb{1}$ is isotropic,
$\bm{W}$ is then a sample from a \emph{Wishart} random matrix ensemble, whose spectral properties are well understood~\cite{mehta}. In the thermodynamic limit of 
infinitely large matrices, $N^{\rm b}\to\infty$, and up to an overall scale of the eigenvalues, the eigenvalue density of $\bm{W}$ is thus given by the
\emph{Mar$\breve{c}$enko-Pastur} law (MP) \cite{pastur} 
\begin{equation}
\label{rho_w}
 \rho_{\alpha}(\hat w)=(1-\alpha)\Theta(1-\alpha)\delta(\hat w)+f_{\alpha}(\hat w),
\end{equation}
where
\begin{equation}
 f_{\alpha}(\hat w)=\frac{1}{2\pi \hat w}\sqrt{(\hat w-\hat w_{-})(\hat w_{+}-\hat w)}
\end{equation}
and is to be read as nonzero only when $\hat w$ lies in the interval 
$[\hat w_{-},\hat w_{+}]$ with $\hat w_{\pm}=\big(\sqrt{\alpha}\pm 1\big)^2$. The delta peak at $\hat w=0$ in \eqref{rho_w} contributes 
only when $\alpha<1$, as indicated by the Heaviside step function $\Theta(\cdot)$. Here we have
defined $\alpha=N^{\rm s}/N^{\rm b}=N^{\text{observed}}/N^{\text{hidden}}$ as the 
fundamental parameter of our analysis, giving the ratio and thus the relative importance of the sizes of the 
observed and unknown ``sectors'' of our network. This parameter resembles the storage ratio \cite{seung, engel}, or number 
of training examples per parameter to be learned, in neural network learning. Indeed, in the context of learning linear relationships
from examples, the distribution \eqref{rho_w} also gives the spectrum of the input correlation matrix governing the learning 
dynamics~\cite{solla,oppersolvable,hertz,sollich}.

In the spectrum \eqref{rho_w} the $\delta$ peak at $\hat w=0$ arises from the $N^{\rm b}-N^{\rm s}=N^{\rm b}(1-\alpha) $ directions in 
the hidden state space that are not directly constrained by observations when $\alpha<1$. 
The remaining $f_{\alpha}(\hat w)$ piece is a semi-circle in the interval $[\hat w_{-},\hat w_{+}]$, distorted by a factor $1/\hat w$. 
For $\alpha>1$ this is the only contribution; in the limit $\alpha\gg1$ the relative variance of the eigenvalues around their mean 
$\langle \hat w\rangle =\alpha$ goes to zero.

\subsection{Self-interacting hidden variables}
\label{sec:selfinteraction}

\subsubsection{Inference error and relaxation times}
\label{sec:justself}

We assume below that the noise acting on bulk variables is isotropic, $\bm{\Sigma}^{\rm bb}=\sigma_{\rm b}^2\mathbb{1}$, as already assumed for the
subnetwork noise. This is equivalent to assuming that the amplitude of fluctuations is homogeneous within the hidden
system, as it would be if it was given by a physical temperature. Anisotropies would add non-trivial correlations between d.o.f.\ 
that would obscure the effect of interactions, 
which is our main focus here. In this section we further restrict ourselves to interactions between bulk and subnetwork, by taking $\bm{K}^{\rm bb}=-\lambda\mathbb{1}$ where the 
self-interaction $\lambda$ is the only interaction among hidden variables. Given this, any interesting behavior has to come from observations.
 
By simultaneously diagonalizing $\bm{W}$ and $\bm{A}$, \eqref{eq:riccati1} reduces to a 
scalar equation relating the eigenvalues of these matrices, respectively $w$ 
and $a$, as 
\begin{equation}
\sigma_{\rm b}^2 a^2 +2\lambda\, a = \frac{k^2}{\sigma_{\rm s}^2}\hat{w},
\end{equation}
where we have extracted from $w$ an amplitude factor by writing $w=k^2\hat{w}/\sigma_{\rm s}^2$, $k$ being the amplitude for the $\ksb$ entries
and $\hat{w}$ a dimensionless Wishart random variable. The physical solution for $a$ is
\begin{equation}
\label{c^F}
 a= \frac{-\lambda+\sqrt{\lambda^2+\sigma^2 \hat{w}}}{\sigma_{\rm b}^2},
\end{equation}
with the shorthand $\sigma=\sigma_{\rm b} k/\sigma_{\rm s}$.
By diagonalizing \eqref{eq:postdrift} one then gets for the eigenvalues of $\kbbs$, which we denote by $r$
\begin{equation}
\label{r}
 \ekbbs=-\lambda -a\,\sigma_{\rm b}^2=-\sqrt{\lambda^2+\sigma^2 \hat{w}}.
\end{equation}
From \eqref{eq:efflineq1} and \eqref{postdecay}, the distribution of $-r$ gives the relaxation rate spectrum of the posterior dynamics, and \eqref{r} shows that these rates are increased by observations, i.e.\ correlations get shorter in time. As expected this effect gets stronger as the hidden-observed interaction amplitude $k$ increases, at fixed
ratio $\sigma_{\rm b}/\sigma_{\rm s}$. 

\begin{figure}
\includegraphics[width=0.48\textwidth]{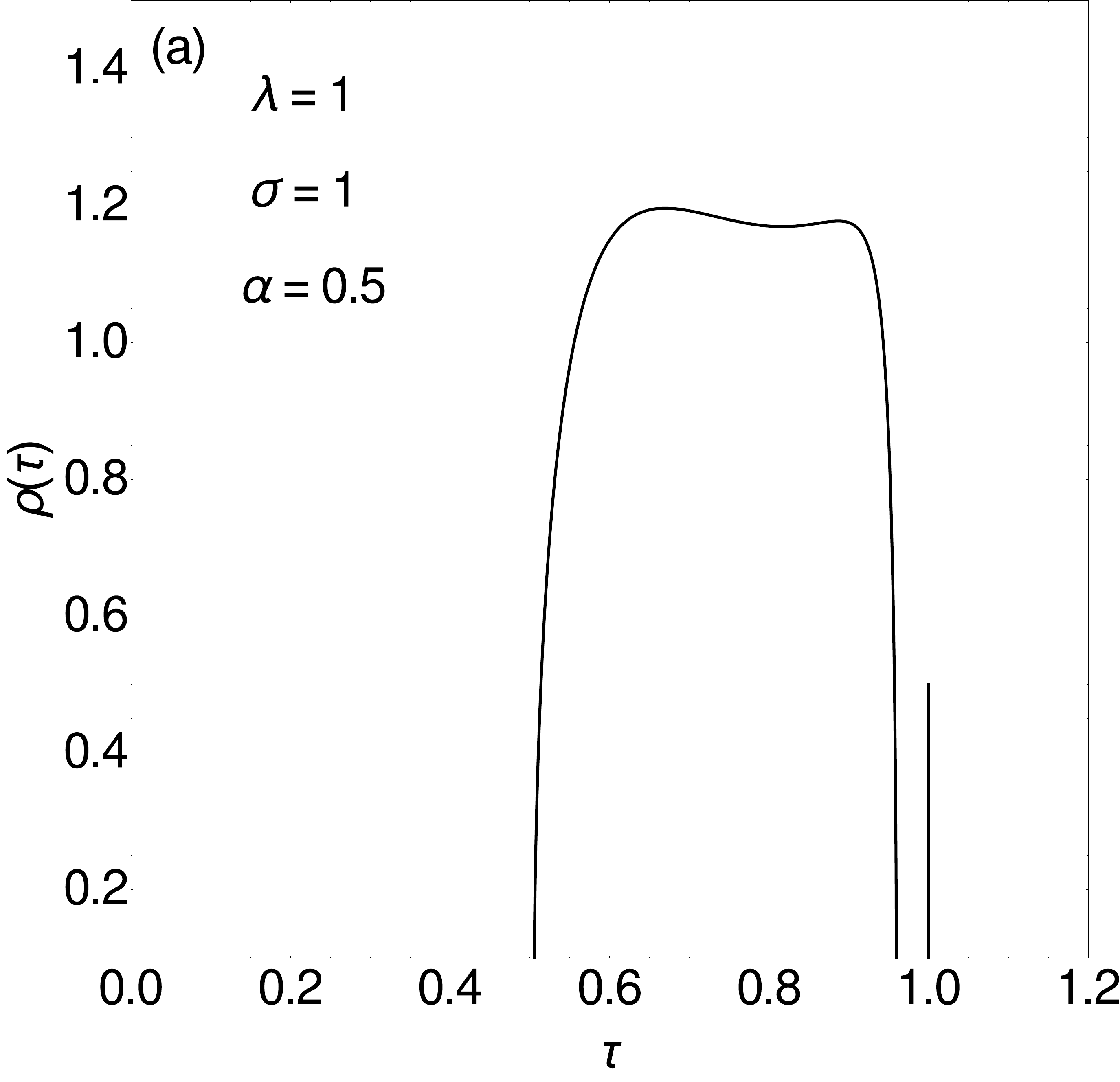}
\includegraphics[width=0.48\textwidth]{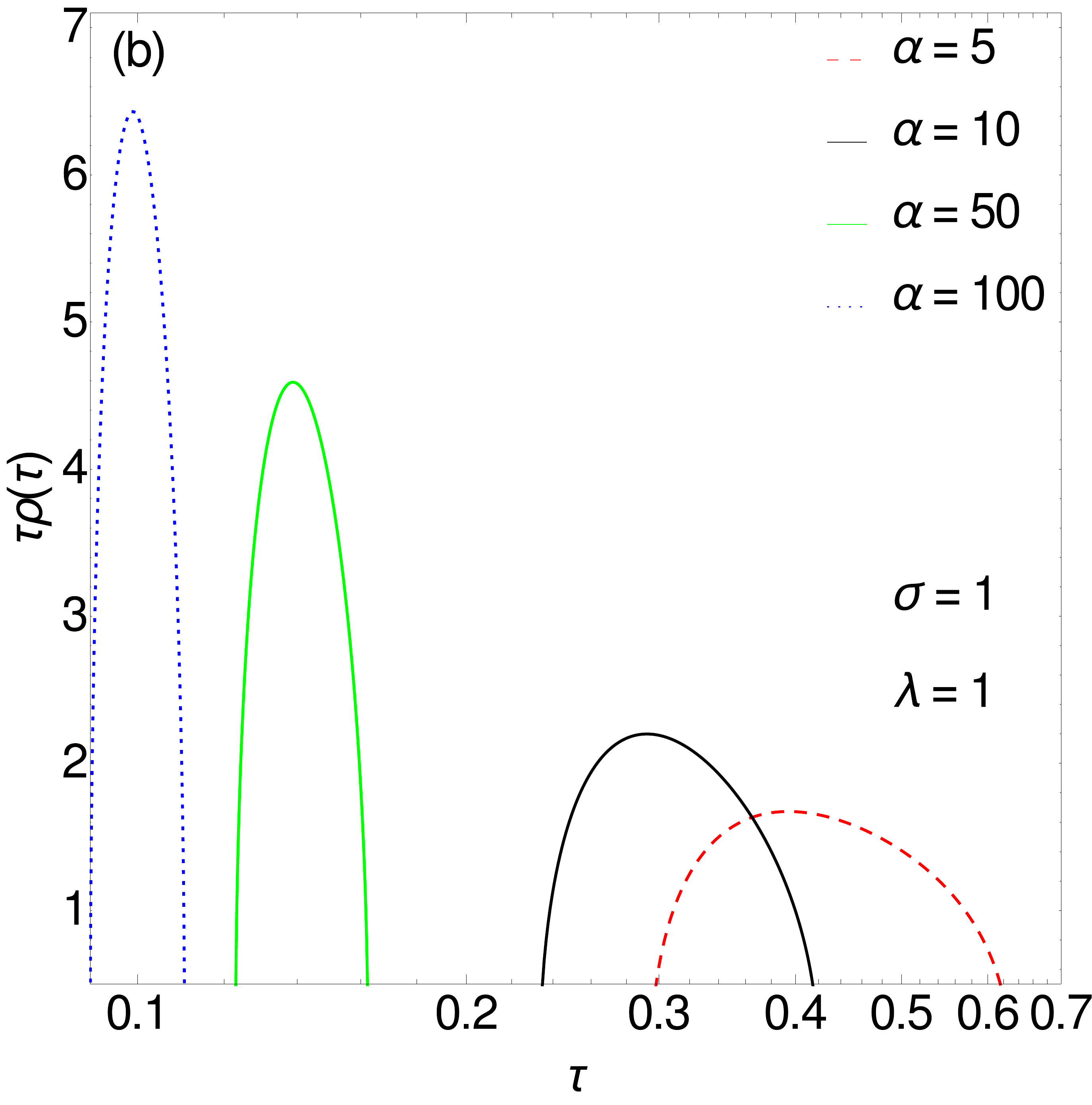}
\caption{(a) Spectral density $\rho(\tau)$ for $\alpha=0.5$: the vertical line indicates the $\delta$-peak of height $1-\alpha$ at $\tau=1/\lambda$, the relaxation time in the absence of observations.
(b) Spectral density $\rho(\ln\tau)=\tau\rho(\tau)$ of $\ln\tau$: this shifts to smaller $\ln\tau$ as  $\alpha$ increases, indicating shorter posterior correlation times. The spectrum also narrows and becomes concentrated around $\tau=1/\sigma\sqrt{\alpha}$ for large $\alpha$.
As the posterior variance $C \propto \tau$ for each hidden space mode, the distributions of $\ln C$ differ only from those of $\ln \tau$ by a horizontal shift.
\label{fig:Only_Self-interactions}
}
\end{figure}

From \eqref{r} we can now find the spectrum of $\ekbbs$ as the appropriate transformation of the MP law 
\begin{equation}
\label{spectime}
 \rho(\ekbbs)=(1-\alpha)\Theta(1-\alpha)\delta(\ekbbs+\lambda)+f(\hat{w}(\ekbbs))|\hat{w}'(\ekbbs)|,
\end{equation}
\normalsize
where $f(\hat{w}(\ekbbs))$ is defined only between $\ekbbs_{\pm}=\sqrt{\sigma^2\big(\sqrt{\alpha}\pm 1\big)^2+\lambda^2}$ and $\hat{w}(r)=-(r^2+\lambda^2)/\sigma^2$ is the inverse function of \eqref{r}. The first piece, a $\delta$-function at $\ekbbs=-\lambda$, describes the behavior for hidden state space directions unconstrained by observations.
The above result for the spectrum can also be expressed as a spectrum $\rho(\tau)=\rho(r)/\tau^2$ of relaxation times 
$\tau=-1/r$ for the posterior dynamics. We sometimes plot $\rho(\ln \tau)=\tau \rho(\tau)$ to show the full range of $\tau$; this $\ln\tau$-spectrum is 
the same as the one of $\ln r$ up to a sign change, with spectral edges at $\tau_{\pm}=-1/\ekbbs_{\mp}$ (see figure \ref{fig:Only_Self-interactions}(a)).

The long-time ($t-t'\gg 1$) behavior of the posterior covariance is an exponential decay whose characteristic time can be defined in different ways. The 
slowest relaxation time is
$\tau_{\text{max}}=1/\ekbbs_{\text{min}}$, where $ \ekbbs_{\text{min}}$ is the minimum eigenvalue of $-\kbbs$
\begin{eqnarray}
\ekbbs_{\text{min}}&=&\sqrt{\lambda^2+\sigma^2 \hat{w}_{\text{min}}}= \nonumber\\
&=&\bigg \{
\begin{array}{rl}
& \lambda\qquad \qquad \qquad \qquad \qquad  \alpha\leq 1\\
&  \sqrt{\lambda^2+\sigma^2(\sqrt{\alpha}-1)^2}\qquad \alpha>1.\\
\end{array}
\end{eqnarray}
One can also look at a relaxation time defined as the average over the spectrum $ \rho(\tau)$,
i.e.\  $\langle \tau \rangle =\int d\tau \rho(\tau)\,\tau$.
Or finally one can consider a root mean square correlation decay time
\be
\label{taustar}
\tau^{*\,2} = \frac{\int_{-\infty}^{+\infty}t^2 C(t)dt}{2\tilde{C}(0)}
=-\frac{1}{2\tilde{C}(0)}\frac{d^2\tilde{C}(\text{i}\omega)}{d^2\omega}\bigg\rvert_{\omega=0},
\ee
where the power spectrum $\tilde{C}(\text{i}\omega)$ is obtained by 
setting $z=\text{i}\omega$ in the Laplace transform (see equation \eqref{osi} below) of the correlator $C(t-t')={\rm Tr}\,\bm{C}(t-t')$ (trace normalized by
$N^{\rm b}$).
It is easy to verify that all three relaxation times exhibit the same asymptotic decay $\sim 1/(\sigma\sqrt{\alpha})$ for large $\alpha$.
In figure \ref{fig:rel-spec}(a) we show a comparison at smaller $\alpha$. With only few observations, all measures of posterior correlation time are close to the $\alpha=0$ value $1/\lambda$ while for $\alpha>1$ they start decreasing, crossing over to the $1/\sqrt{\alpha}$ large $\alpha$ tail; $\tau_{\text{max}}$ shows
the least smooth transition between these two regimes. We can summarize the behavior by saying that with more observations the posterior fluctuations (or error bars on the inferred means) become less correlated in time as predictions become more ``tied'' to the data observed at any specific moment.
This effect is seen in more detail in figure \ref{fig:Only_Self-interactions}(b) where with increasing $\alpha$ the relaxation time spectrum becomes more peaked and shifts towards shorter times.
The posterior covariance matrix $\bm{C}$ has the same set of eigenmodes as $\bm{K}^{\rm bb|s}$ in the current scenario because in \eqref{eq:post} 
all matrices can be simultaneously diagonalized. The eigenvalues $C$ of $\bm{C}$ give the posterior variance for each mode, which 
from \eqref{eq:post} is related to $r$ or $\tau$ by
\begin{equation}
\label{eq:C}
 C=-\frac{\sigma_{\rm b}^2}{2 \ekbbs}=
\frac{\sigma_{\rm b}^2}{2}\tau =  
 \frac{\sigma_{\rm b}^2}{2\sqrt{\lambda^2+\sigma^2 \hat{w}}}.
\end{equation}
This shows that $C$ decreases with increasing feedback values $\hat{w}$:  
observations increase prediction accuracy as they should. Because $C\propto\tau$, the above results for the spectrum of $\tau$ also apply to that of $C$; see 
figures \ref{fig:Only_Self-interactions} and \ref{fig:rel-spec}(a). 
For large $\alpha$ in particular the spectrum of $C$ becomes a narrow peak around the asymptotic inference error $C\approx \sigma_{\rm b}^2/(\sigma\sqrt{\alpha})$.

We note as an aside that from the proportionality $C\propto \tau$ one can show that the relaxation time $\tau^*$ defined in \eqref{taustar} can be written in terms of spectral averages as
\be
\tau^*= \sqrt{\frac{\langle \tau^4 \rangle}{\langle \tau^2 \rangle}}.
\ee
Because $\langle \tau \rangle^2 \langle \tau^2 \rangle \leq \langle \tau^4 \rangle$, this implies generally $\langle \tau \rangle \leq \tau^*$ in agreement with the results in figure \ref{fig:rel-spec}(a).

\subsubsection{Posterior covariance in Laplace space}
\label{sec:post_var_wis_LS}
We next turn to the temporal dependence of the posterior covariance \eqref{postdecay}. Its trace, 
normalized by $N^{\rm b}$, is an average of the contributions from the different eigenmodes of $\kbbs$. 
In terms of the relevant eigenvalues $\hat{w}$ and using \eqref{eq:C} these are 
\begin{equation}
\label{eq:C(t)}
C_{\hat{w}}(t-t')= e^{\ekbbs|t-t'|}C= -\frac{\sigma_{\rm b}^2}{2 \ekbbs}e^{\ekbbs|t-t'|},
\end{equation}
with an added subscript ${\hat{w}}$ to indicate this is the contribution from a single eigenmode, characterized by a specific value of ${\hat{w}}$.
We take the double-sided Laplace transform
\begin{eqnarray}
\label{eq:dsL}
\tilde{C}_{\hat{w}}(z)&&=
\frac{\sigma_{\rm b}^2}{2r}\int_{-\infty}^{+\infty}e^{-(z+\ekbbs)|t'-t|}dt'\notag\\
&&=\frac{\sigma_{\rm s}^2}{k^2} \frac{1}{\frac{\lambda^2-z^2}{\sigma^2}+ \hat{w}},
\end{eqnarray}
where we have substituted \eqref{r} for $\ekbbs$ in terms of
the self-interaction $\lambda$ and the feedback matrix eigenvalues $k^2\hat{w}/\sigma_{\rm s}^2$. 

In the thermodynamic limit, we can then get the Laplace transform of the overall covariance normalized trace $C(t-t')={\rm Tr}\,\bm{C}(t-t')$ 
by averaging over the Mar$\breve{\text{c}}$enko-Pastur spectrum $\rho(\hat{w})$, yielding
\begin{widetext}
\begin{equation}
\label{osi}
\tilde{C}(z)=\frac{\sigma_{\rm s}^2}{2 k^2}\frac{\sigma^2}{{(\lambda^2-z^2)}}\bigg\lbrace 1-\alpha-\bigg(\frac{\lambda^2-z^2}{\sigma^2}\bigg)+
\sqrt{\bigg[1-\alpha- \bigg(\frac{\lambda^2-z^2}{\sigma^2}\bigg)\bigg]^2+4\bigg(\frac{\lambda^2-z^2}{\sigma^2}\bigg) }\bigg\rbrace.
\end{equation}
\end{widetext}

\begin{figure}
\includegraphics[width=0.48\textwidth]{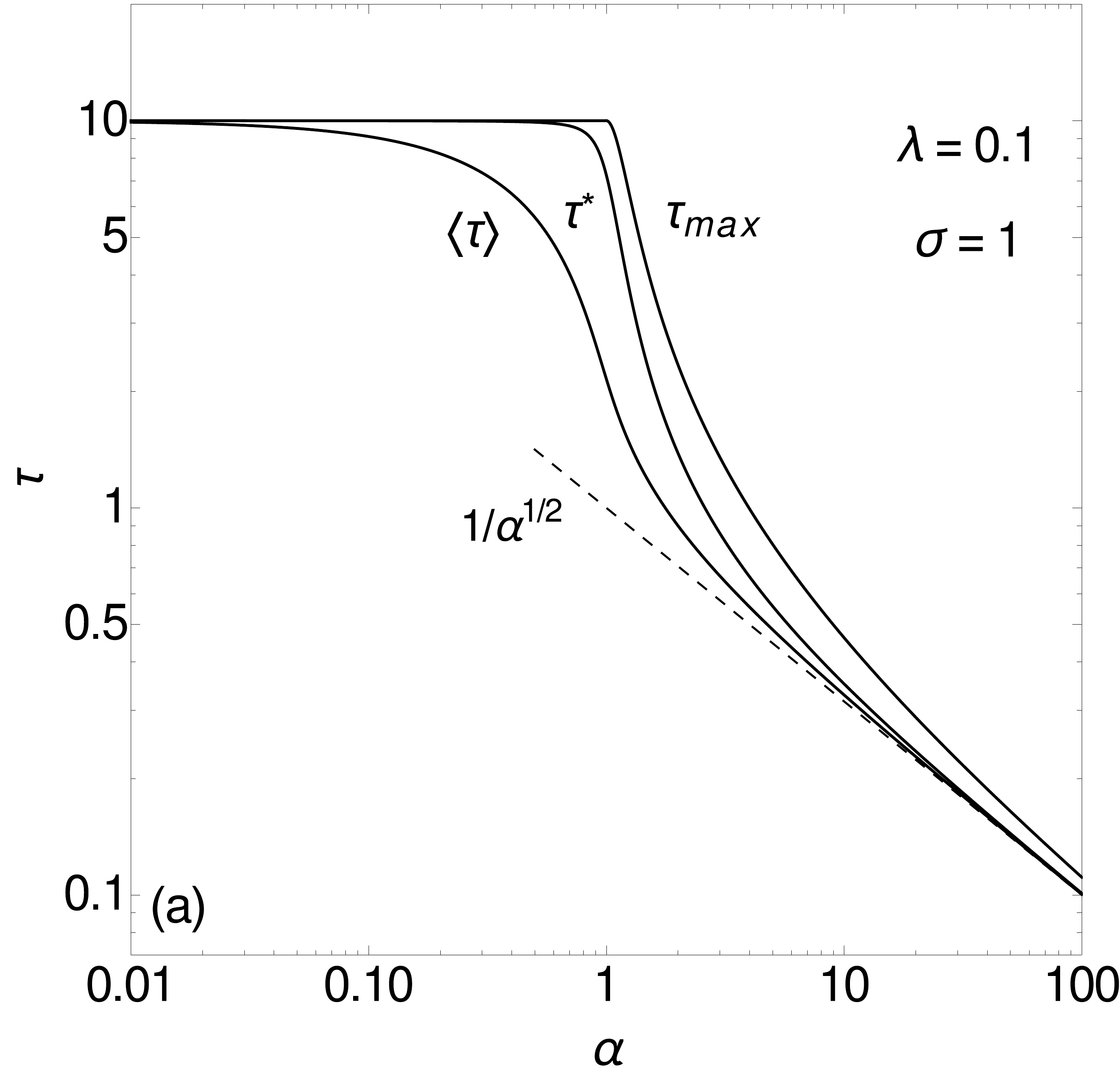}
\includegraphics[width=0.48\textwidth]{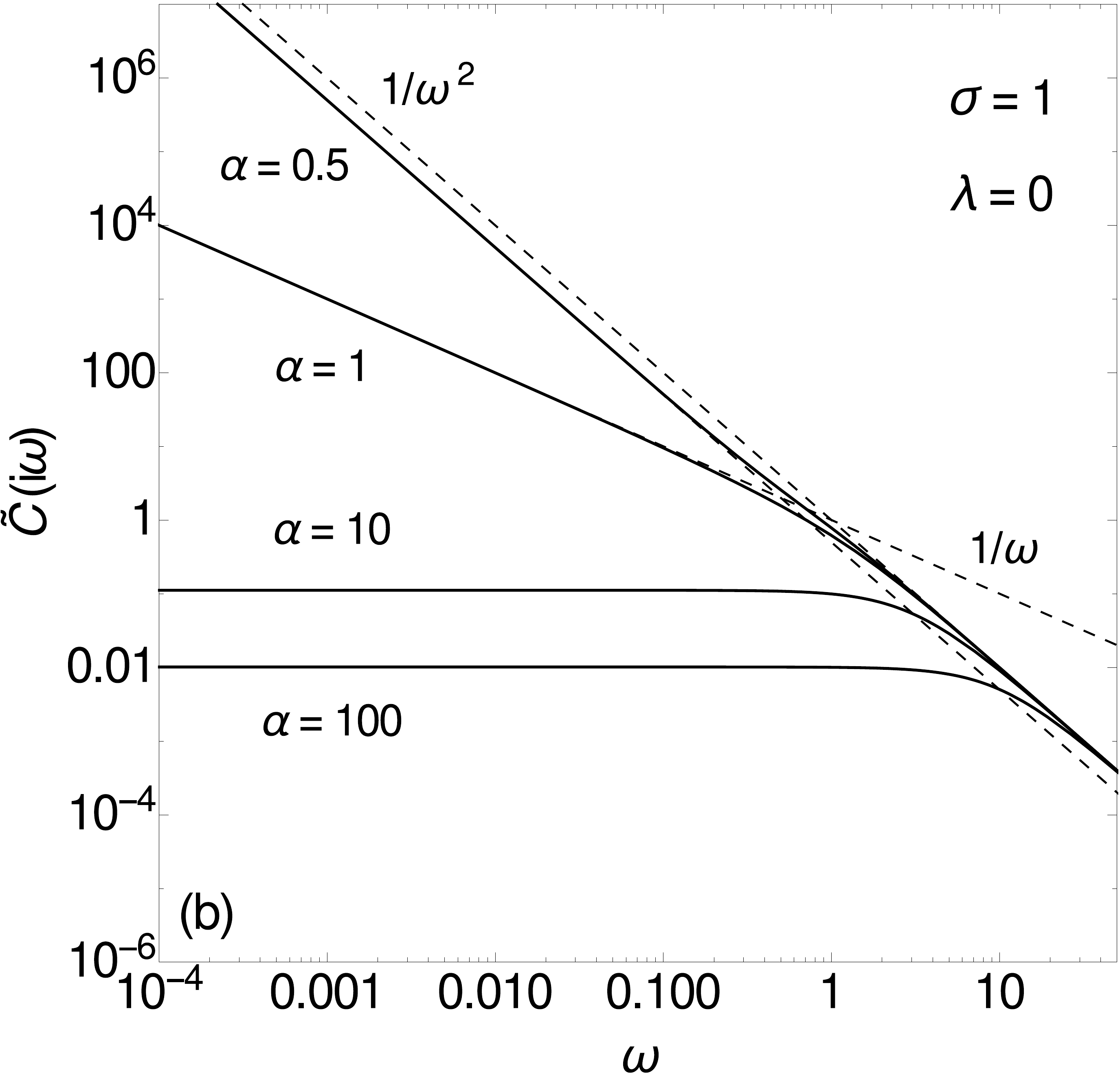}
\caption{(a) Characteristic posterior relaxation time $\tau$ as a function of $\alpha$, for $\lambda=0.1$ and $\sigma=1$, defined in three different ways (see text). For $\alpha \to 0$ all three curves approach $\tau=1/\lambda=10$; asymptotically they decay as 
$1/\sqrt{\alpha}$. 
(b) Posterior power spectrum (obtained by 
setting $z=\text{i}\omega$ in \eqref{osi}) for various $\alpha$, at $\lambda = 0$. The power spectrum diverges as $\omega\to 0$ when $\alpha\leq 1$. 
For small $\alpha$ the divergence is  $\propto 1/\omega^2$, crossing over to $\propto 1/\omega$ as $\alpha\to 1$.
Beyond $\omega\sim O(1)$ the curves for all $\alpha$ exhibit a standard Lorentzian tail $1/\omega^2$. See \cite{plefkaobs2}
for a derivation of these power laws.} 
\label{fig:rel-spec}
\end{figure}

One can verify that $\tilde{C}(0)$ has a divergence for 
$\lambda/\sigma\to0$ and $\alpha \leq 1$; the small $\alpha$-curves in figure \ref{fig:rel-spec}(b) illustrate this effect. 
See also \cite{plefkaobs2} for a systematic study of the approach to such divergences.

\subsection{Symmetric hidden-hidden couplings}
\label{sec:symmetric_couplings}

In this section we generalize the above scenario 
by assuming that  $\jm^{\rm bb}=-\lambda\mathbb{1}+\bm{J}$. Here the matrix $\bm{J}$ provides explicit hidden-to-hidden interactions beyond the self-interaction term $-\lambda \mathbb{1}$ we have had so far. 
To ensure stability of the hidden system, one requires $\lambda > \lambda_{\rm c}$ where  
 $\lambda_{\rm c}$ is the largest eigenvalue of $\bm{J}$. 

We assume that $\bm{J}$ is symmetric, which is required for any steady state of the whole system to be at equilibrium, i.e.\ to obey detailed balance.
The posterior drift $\jm^{\rm bb|s}$ 
from \eqref{eq:postdrift} is then also a symmetric matrix.
This is crucial as it allows one to solve \eqref{eq:post} and \eqref{eq:riccati1} in closed form. Eq.~\eqref{eq:post} gives 
\begin{equation}
 \bm{C} =
 -\frac{\sigma_{\rm b}^2}{2} \big(\jm^{\rm bb|s}\big)^{-1},
\end{equation}
which is positive definite because $\jm^{\rm bb|s}=(-\lambda+\bm{J})-\sigma_{\rm b}^2\bm{A}$ is negative definite. To eliminate the unknown $\bm{A}$, note from \eqref{eq:riccati1} that
\begin{equation}
\begin{split}
&\bigg((-\lambda+\bm{J}) - \sigma_{\rm b}^2\bm{A}\bigg)^2=
(-\lambda+\bm{J})^2+\sigma_{\rm b}^4\bm{A}^2\\
&-\sigma_{\rm b}^2(-\lambda+\bm{J})\bm{A}-
\bm{A}(-\lambda+\bm{J})\sigma_{\rm b}^2=\\
&=(-\lambda+\bm{J})^2+\sigma_{\rm b}^2\bm{W} \doteq \bm{M},
\end{split}
\end{equation}
where the last equality defines $\bm{M}$. Hence
\begin{equation}
\label{eq:candm}
  \bm{C}
 =\frac{\sigma_{\rm b}^2}{2} \bm{M}^{-1/2},
 \qquad
 \jm^{\rm bb|s} = - \bm{M}^{1/2},
\end{equation}
where $\bm{M}^{1/2}$ is the positive definite square root of $\bm{M}$ and $\bm{M}^{-1/2}$ its inverse.

\subsubsection{Free probability}

From \eqref{eq:candm}, the spectrum of $\bm{M}$ directly determines those of $\bm{C}$ and $\jm^{\rm bb|s}$. As a paradigmatic example where this spectrum can be obtained in the thermodynamic 
limit we consider the case where the elements of $\bm{J}$ are independently drawn from a Gaussian distribution, i.e.\ we set $\bm{J}=j\bm{\hat{J}}$ with
$\bm{\hat{J}}$ a random matrix from the  \emph{Wigner} ensemble~\cite{mehta}. 
From the Wigner semi-circular law this has largest eigenvalue 2, thus $\lambda_{\rm c}=2j$. 
We will write the feedback matrix  as in 
section \ref{sec:post_var_wis_LS}: $\bm{W}=\frac{k^2}{\sigma_{\rm s}^2}\bm{\hat{W}}$ with $\bm{\hat{W}}$ from the \emph{Wishart} ensemble.

With the above assumptions,
$\bm{M}=(-\lambda+\bm{J})^2 +\sigma_{\rm b}^2\bm{W}$ is a sum of two independently drawn, symmetric random matrices with known spectrum.
Its spectrum can then be found using \emph{free probability} theory. Reviews can be found in~\cite{voiculescu} for the theory and \cite{speicher,burda} for applications to RMT.
Briefly, the sum defining $\bm{M}$ is effectively a \emph{free} addition \cite{voiculescu} in the sense that because of independent sampling, the eigenvector bases of the two matrices in the sum are randomly rotated against each other.
It then turns out that the spectrum of the sum depends only on the eigenvalues and not the eigenvectors of the individual matrices.
The intuition beyond this
is that, in the limit of infinite matrix size, the detailed statistics of eigenvalues, e.g.\ whether they are correlated or not, can be neglected \cite{burda}. 
While in an ordinary sum of independent random variables it is the cumulants that add, in a free sum of two random matrices 
it is the $R$-transforms that are additive \cite{voiculescu}, and this allows the spectrum of the sum to be determined.

The $R$ transform of a random matrix is related to its Green's function by
\begin{equation}
\label{RG}
 G(z)=\frac{1}{z-R(G(z))}.
\end{equation}
The  Green's function or resolvent, in turn, is defined for a generic random matrix $\bm{M}$ as the normalized trace 
$G_M(z)=\text{Tr}(z-\bm{M})^{-1}$. It can be written in terms of the eigenvalue density $\rho(m)$ as 
\be
G_M(z) = \int \frac{\rho(m)}{z-m}dm,
\ee
which is also known as a Stieltjes transform. Conversely, $\rho(m)$ can be retrieved from the Green's function via
\begin{equation}
\label{eq:grho}
 \rho(m)=-\frac{1}{\pi}\lim_{\epsilon\rightarrow 0^+}\text{Im}\,G_M (m+\text{i}\epsilon).
\end{equation}
The route to finding the spectrum of $\bm{M}$ in our case is then clear: 
we need to write the Green's functions and associated $R$-transforms of $(-\lambda+\bm{J})^2$ and $\sigma_{\rm b}^2\bm{W}$, 
respectively, add these two $R$-transforms to obtain the $R$-transform of $\bm{M}$, and then work backwards to $G_M(z)$ and finally $\rho(m)$.

We denote by $G_1(z)$ the Green's function of $(-\lambda+\bm{J})^2$, which is given by the integral
\begin{eqnarray}
\label{eq:green1}
 G_1(z)&=&\int \frac{\rho(\hat{\jmath})}{z-(-\lambda+j\hat{\jmath})^2}d\hat{\jmath}\nonumber\\
 &=&\int_{-2}^{2} \frac{\sqrt{4-\hat{\jmath}^2}}{2\pi}\frac{1}{z-(-\lambda+j\hat{\jmath})^2}d\hat{\jmath},
\end{eqnarray}
where the Wigner semicircular law has been used.
The integral can be performed in closed form
\begin{eqnarray}
\label{eq:grsym}
 G_1(z)&=&\frac{1}{2 j^2}-\frac{1}{4j^2}\sqrt{\frac{\big(\lambda-\sqrt{z}\big)^2-4 j^2}{z}}\nonumber\\
 &&{}-\frac{1}{4j^2}\sqrt{\frac{\big(\lambda +\sqrt{z})^2-4 j^2 }{z}}
\end{eqnarray}
and  \eqref{RG} then gives the $R$-transform \begin{equation}
 R_1(z)=\frac{j^2}{1-z j^2}+\frac{\lambda^2}{\big(1-2 z j^2 \big)^2}.
\end{equation}
The Green's function for a Wishart matrix is well known \cite{hertz} and the related $R$ transform reads
\begin{equation}
 R_2(z)=\frac{\alpha v}{1-v z},
\end{equation}
where we recall that $\alpha=N^{\rm s}/N^{\rm b}$ and $v$, the variance, in our case is $v=k^2\sigma_{\rm b}^2/\sigma_{\rm s}^2$.
The two above $R$-transforms now simply add to give the one for $\bm{M}$, ${R}_M(z) = R_1(z) + R_2(z)$.
The result can be written as an implicit expression for the Green's function $G_M(z)$, given that from \eqref{RG} one has generally $z(G) = 1/G + R(G)$
\begin{equation}
\label{greensym}
 z=\frac{1}{G}+\frac{\alpha \frac{k^2\sigma_{\rm b}^2}{\sigma_{\rm s}^2}}{1-\frac{k^2\sigma_{\rm b}^2}{\sigma_{\rm s}^2}G}
 +\frac{j^2}{1-j^2G}+\frac{\lambda^2}{\big(1-2 j^2 G\big)^2}.
\end{equation}
We have abbreviated $G\equiv G_M$ on the r.h.s.\ here. Rearranging 
the above equation one sees that $G(z)$ is the solution of a fifth order polynomial equation. 
This can be found numerically, with the correct solution branch being determined from the asymptotic behavior $G\approx 1/z$ for large $z$. 
Once $G(z)$ is in hand, $\rho(m)$ can be found using \eqref{eq:grho}. 

\begin{figure}
\includegraphics[width=0.48\textwidth]{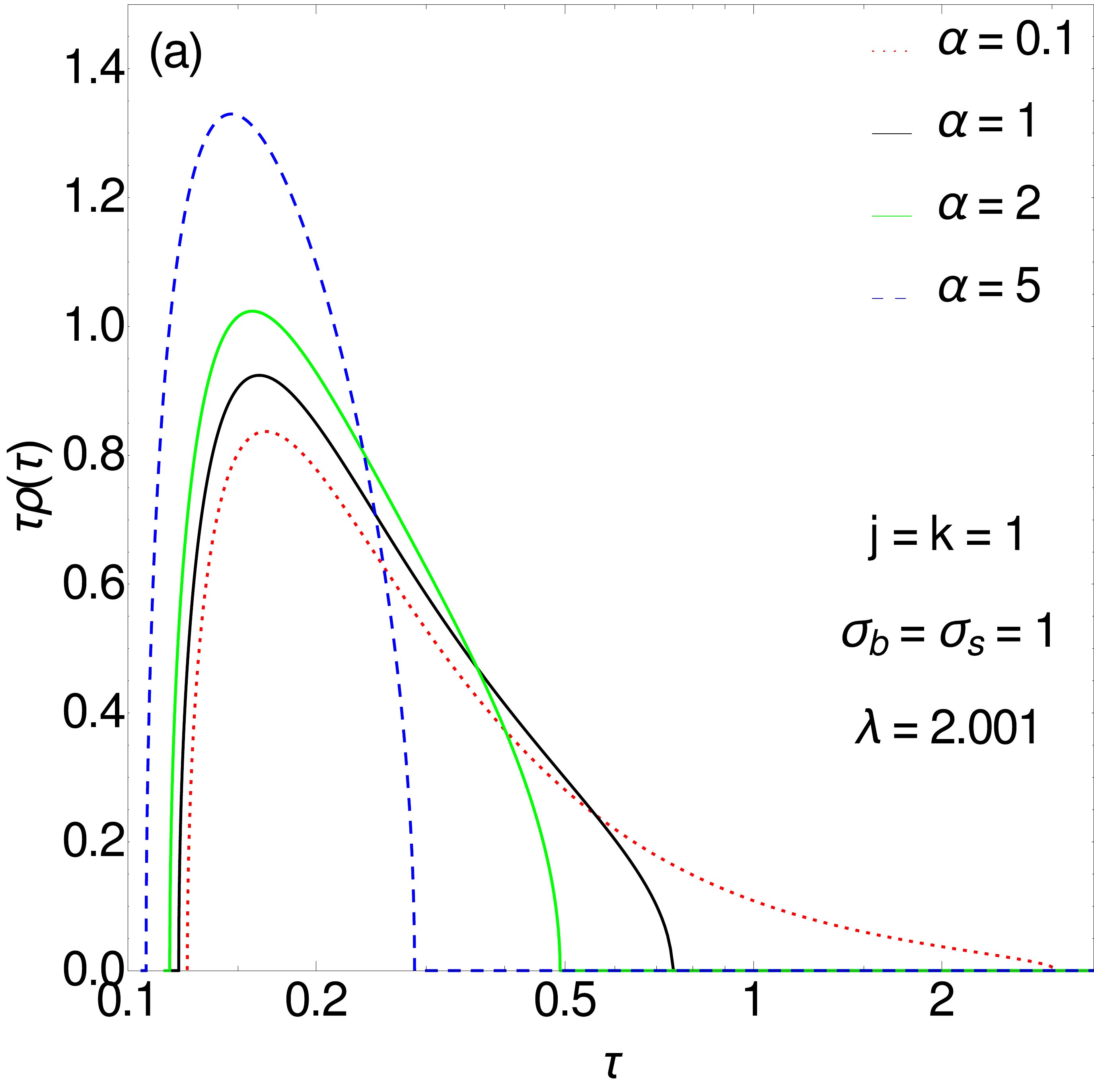}
\includegraphics[width=0.48\textwidth]{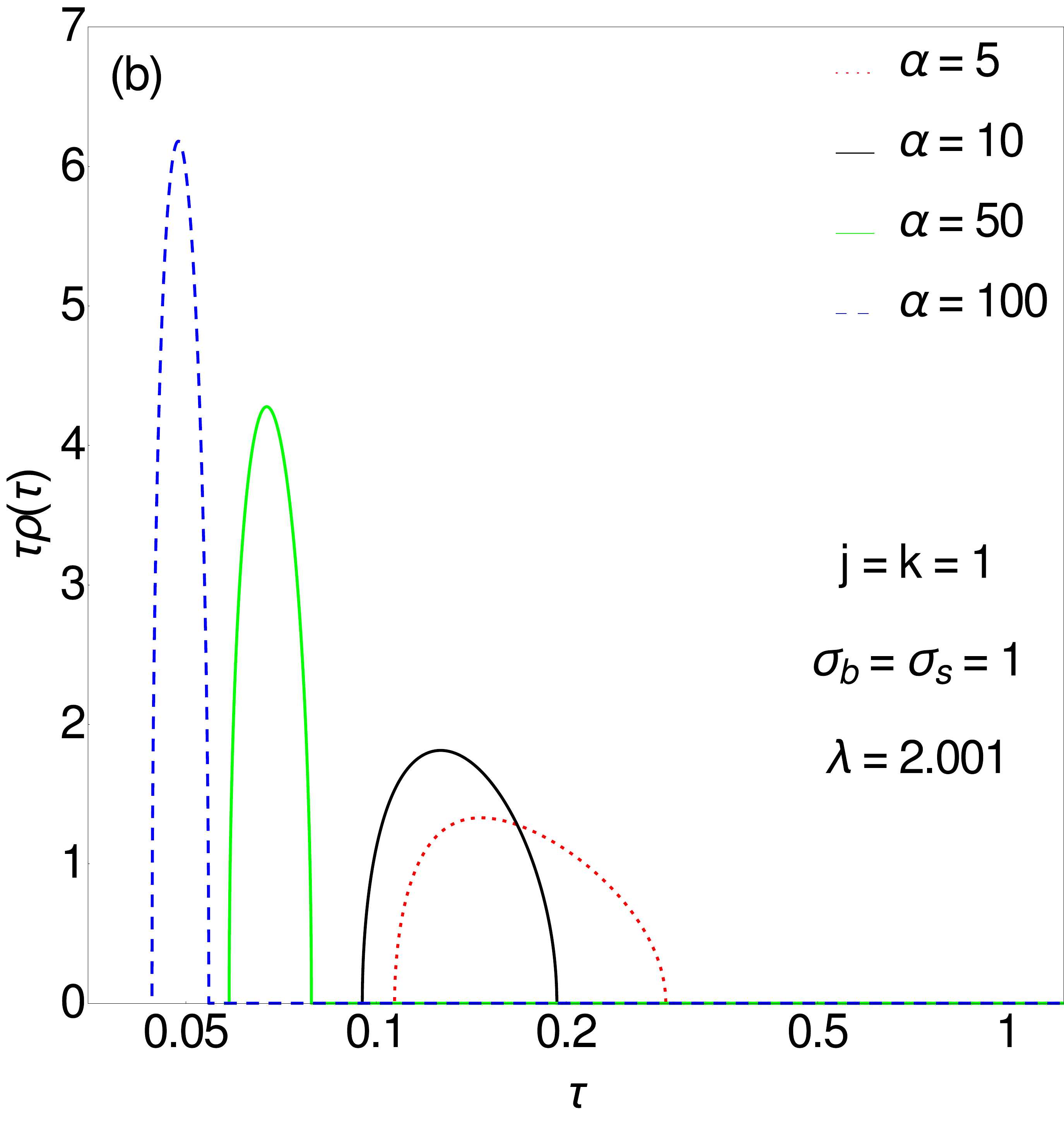}
\caption{Spectral density $\rho(\ln\tau)=\tau\rho(\tau)$, of relaxation times $\tau$, for different values of 
$\alpha$. We plot $\rho(\ln\tau)$ to make the normalization of the densities more obvious. The spectra of posterior variances $C$, which define the inference error, are identical up to a horizontal shift as $C\propto \tau$. 
(a) At small $\alpha$ the spectrum is broad, indicating that there is much variation in how different hidden state space directions are constrained by observations. For
increasing $\alpha$ the spectrum becomes more peaked, and centred around decreasing $\tau$ or $C$: different directions become determined more strongly, and more evenly, by observations, a trend more clearly visible in (b).}
\label{fig:rho_symmetric}
\end{figure}

By a transformation of the spectrum of $\bm{M}$ we can characterize the spectrum of the posterior covariance matrix
$\bm{C}=\sigma_{\rm b}^2\bm{M}^{-{1}/{2}}/2$ as well as the spectrum of relaxation rates as determined by the effective drift
$\kbbs=-\bm{M}^{{1}/{2}}$. The spectrum of $(-\kbbs)^{-1}=\bm{M}^{-1/2}$ then gives the distribution of relaxation times. As this matrix is proportional to $\bm{C}$, 
plots of $\rho(\tau)$ (figure \ref{fig:rho_symmetric}) provide information also about the inference error as a function of $\alpha$. The overall picture is that predictions become increasingly precise when the pool of observed data is expanded, i.e.\ $\alpha$ increases, while correlation times between posterior fluctuations decrease in proportion.
 
For qualitative analysis  one can rewrite \eqref{greensym} in dimensionless variables 
$\tilde{z}=\sigma_{\rm s}^2z/(k^2 \sigma_{\rm b}^2)$ and 
$\tilde{G}= k^2 \sigma_{\rm b}^2G/\sigma_{\rm s}^2$ as
\begin{equation}
\label{greensymadim}
 \tilde{z}=\frac{1}{\tilde{G}}+\frac{\alpha}{1-\tilde{G}}+\frac{(\gamma p)^2}{1-(\gamma p)^2\tilde{G}}+
 \frac{p^2}{\big(1-2(\gamma p)^2\tilde{G}\big)^2},
\end{equation}
where $\gamma= j/\lambda$ and $p=\lambda/\sigma$. This reduces the number of parameters and variables, from seven 
($\alpha$, $j$, $k$, $\lambda$, $\sigma_{\rm s}$, $\sigma_{\rm b}$, $z$) to four ($p$, $\gamma$, $\alpha$, $\tilde{z}$). 
Here $\gamma$ and $1/p$ measure the strength of hidden-hidden and hidden-observed couplings relative to the decay weight $\lambda$.

\begin{figure}
\includegraphics[width=0.49\textwidth]{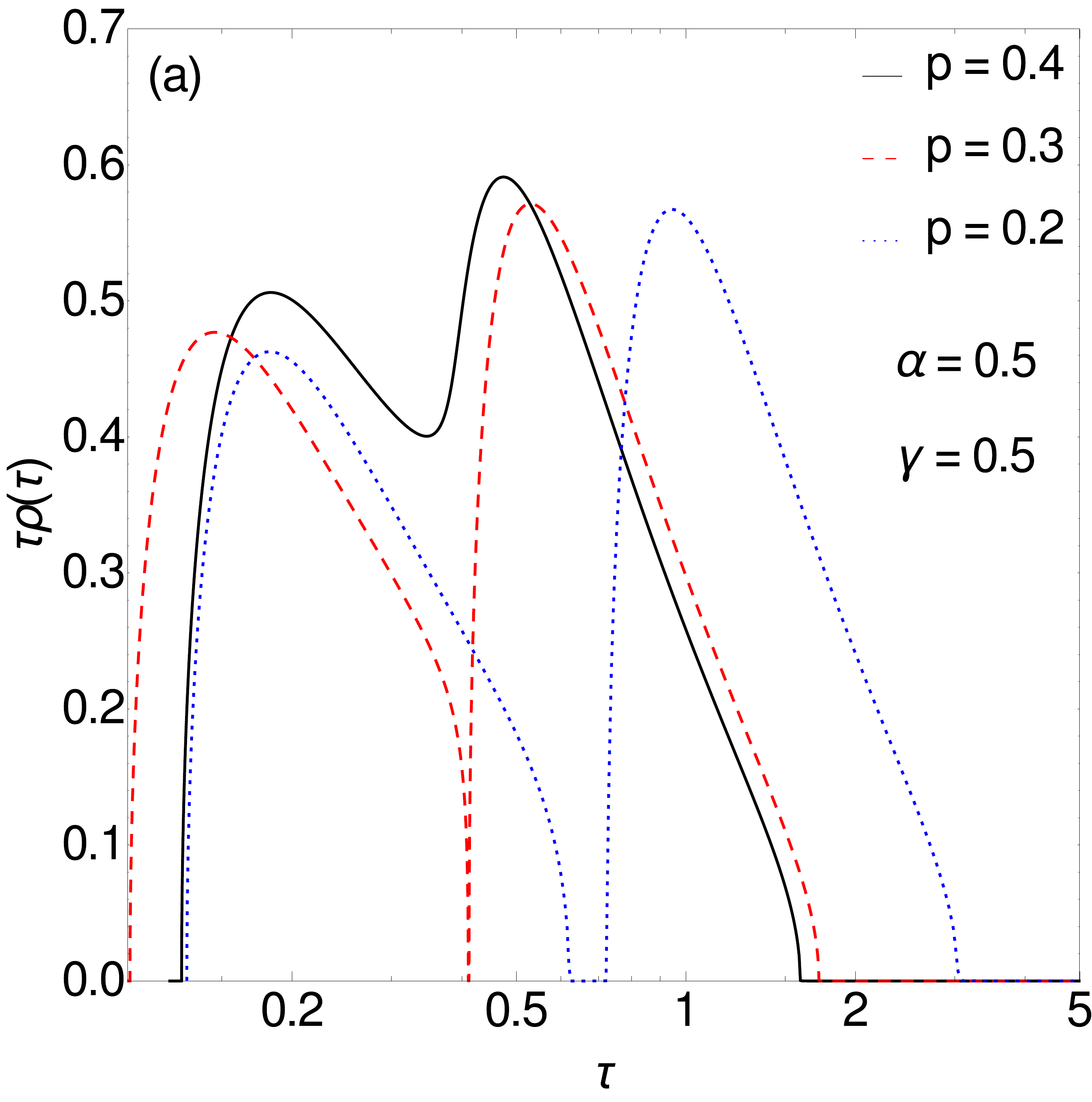}
\includegraphics[width=0.5\textwidth]{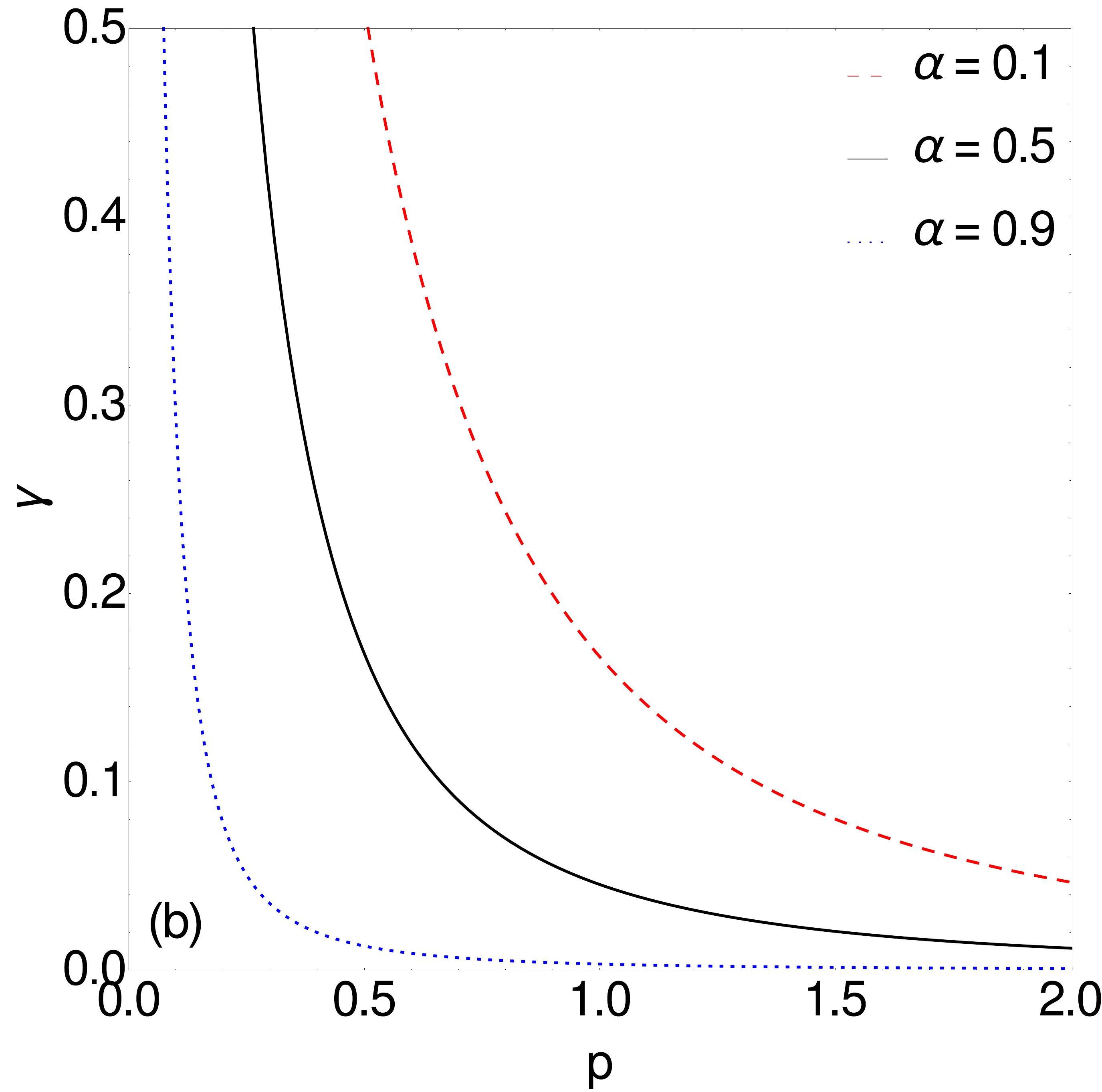}
\caption{(a) Spectral density $\rho(\ln\tau)=\tau\rho(\tau)$, at $\gamma=j/\lambda=0.5$ (critical value for internal stability, with $j=0.2$ and $\lambda=0.4$) and $\alpha=0.5$ 
for different values of $p$: the two pieces of the spectrum at $p=0.2$  merge at $p=0.3$, giving a spectrum supported on a single interval for $p>0.3$. 
(b) Curve in the $(p,\gamma)$ plane for which the two pieces of the spectrum merge when coming from low $\gamma$: the black line
refers to $\alpha=0.5$, the case shown in (a). The two-piece region near the origin shrinks (see curve for $\alpha=0.9$, blue dotted line) and vanishes for $\alpha\to 1$. }
\label{fig:rho_alpha}
\end{figure}

We have seen in figure \ref{fig:Only_Self-interactions}(a) that for $\gamma=0$, i.e.\ in the absence of hidden-hidden interactions (see section \ref{sec:justself}) the spectrum consists of two separate pieces for $\alpha<1$, while with such interactions present ($\gamma>0$) the spectrum can be supported on a single interval. 
There must be a transition between these two cases at some value of $\gamma$ that will depend on $p$ and $\alpha$ - see figure \ref{fig:rho_alpha} 
(a). Locating this transition numerically gives the results shown in figure \ref{fig:rho_alpha}(b). The spectrum consists of a single piece {\em above} the line drawn in the $(p,\gamma)$ plane. One sees that for 
large $p=\lambda/\sigma =  \lambda\sigma_{\rm s} /(\sigma_{\rm b} k)$, i.e.\ weaker hidden-observed couplings, small values of $\gamma=j/\lambda$ and 
hence weak hidden-hidden interactions are sufficient to merge the two pieces of the spectrum.

\subsubsection{Posterior correlations in Laplace space}
\label{sec:post_symm}
From \eqref{postdecay} and \eqref{eq:candm} 
we can obtain explicitly the posterior correlations in time: for $t>t'$,
 \begin{equation}
  \bm{C}(t-t')=\frac{\sigma_{\rm b}^2}{2}e^{-\bm{M}^{1/2}(t-t')}\bm{M}^{-1/2}.
 \end{equation}
We consider the trace, which at $t=t'$ gives the total posterior variance. The double-sided Laplace transform can then be shown to have the simple form
\begin{equation}
\label{ctog}
 \tilde{C}(z)=\sigma_{\rm b}^2\,\text{Tr}\left(-z^2+\bm{M}\right)^{-1} =-\sigma_{\rm b}^2 G_M(z^2).
 \end{equation}
This relation to the Green's function is in fact a statement of the Fluctuation-Dissipation Theorem \cite{FDT} (see \cite{thesis} for details) and holds true because of the symmetry of $\bm{J}$.
 
From \eqref{ctog}, the Laplace transformed posterior correlation function has to 
satisfy the equation for $-\sigma_{\rm b}^2 G_M(z^2)$, giving
 \begin{equation}
 \label{algcorr}
  z^2=-\frac{\sigma_{\rm b}^2}{\tilde{C}}+\frac{\alpha\frac{k^2 \sigma_{\rm b}^2}{\sigma_{\rm s}^2}}{1+\frac{k^2}{\sigma_{\rm s}^2}\tilde{C}}+
  \frac{j^2}{1+\frac{j^2}{\sigma_{\rm b}^2}\tilde{C}}+\frac{\lambda^2}{\bigg(1+2\frac{j^2}{\sigma_{\rm b}^2}\tilde{C}\bigg)^2},
 \end{equation}
where we have set $\tilde{C}(z)=\tilde{C}$. Interestingly, and similarly to \eqref{greensym} which determines the spectrum of $\bm{M}$, this 
equation does not become singular at $\lambda=0$.
This fact can be understood in the following way. 
If directions exist along which the hidden dynamics would grow exponentially without observations, then these always have a non-zero overlap with directions constrained by observed data. This is clear from the independent sampling of the two terms in $\bm{M}$, and explains how the posterior variance, the uncertainty on the hidden dynamics, can stay finite even when the hidden dynamics without observations would diverge.
Nevertheless, such a diverging hidden dynamics is an unphysical situation. We therefore continue to consider only parameter sets with $\lambda> \lambda_{\rm c}$, the internal dynamical condition for a finite and well-defined marginal dynamics of the bulk. 

Finally, by setting $z=\text{i}\omega$ one can evaluate the posterior power spectrum $\tilde{C}(\text{i}\omega)$.  
It can be written in terms of a dimensionless function $\ac_{\alpha,p,\gamma}(\omr)$
 \begin{equation}
  \tilde{C}(\text{i}\omega)=\frac{\sigma_{\rm s}^2}{k^2}\ac_{\alpha,p,\gamma}\big(\omr\big),
 \end{equation}
with $\omr=\omega/\sigma$ a rescaled frequency. The prefactor shows that the entire power spectrum of the posterior variance or prediction uncertainty is directly proportional to 
the dynamical noise acting on the observed subnetwork $\sigma_{\rm s}^2$
and inversely proportional to $k^2$, the strength with which it interacts with the bulk. 
As before one can find from \eqref{algcorr} an equation for the dimensionless part $\ac$
\begin{equation}
\label{eq:adimsym}
 -\omr^2 = -\frac{1}{\ac}+\frac{\alpha}{1+\ac}+\frac{(\gamma p)^2}{1+(\gamma p)^2 \ac}+\frac{p^2}{\big(1+2(\gamma p)^2 \ac \big)^2},
\end{equation}
where $\gamma$ and $p$ are defined as before.
One can verify that for $p=0$ and $0\leq \alpha \leq 1$, $\ac(0)$ has a divergence, implying also that the time integral of $\text{Tr}\,\bm{C}(t-t')$ diverges. This comes physically from the fact that while a fraction $\alpha$ of hidden space directions have variances (and co-variances) of the expected order $\propto 1/k^2$, the others have variances that are independent of $k$ and therefore much larger for large $k$.

A second region in the $\alpha$, $p$, $\gamma$
parameter space where $\ac(0)$ diverges is $\alpha\to 0$ and $\gamma \to \gamma_c=1/2$. This is as expected: without observations, the hidden dynamics starts to diverge at 
$\lambda \to \lambda_{\rm c}=2j$, hence at $\gamma_c=1/2$. We refer 
to \cite{plefkaobs2} for further discussion
of the behavior in the vicinity of such critical points.

The results of this section are of conceptual and practical significance. First, equation \eqref{greensym}
for the Green's function provides a tool to study in a controlled way how spectra change with the number
of observations and the interaction strength: this is what we show in figures  \ref{fig:Only_Self-interactions}, \ref{fig:rho_symmetric} and \ref{fig:rho_alpha}. Second, as more
thoroughly analyzed in \cite{plefkaobs2}, from equations \eqref{algcorr} and \eqref{eq:adimsym} one can calculate posterior equal time variances
(by Fourier Transform) and relaxation times (by the
second derivative at zero frequency, see \eqref{taustar}), which are
exact in the thermodynamic limit and thus expected
to be good approximations for large size datasets.
Importantly, exact values such these can serve as
a reference point around which one could systematically investigate finite size effects.

\section{Thermodynamic Limit by Dynamical Functionals}
\label{Dynamical_functional}

So far we have studied the posterior variance and time-dependent covariance in settings where the dynamics of the entire network obeys detailed balance, 
and where the relevant Green's functions can be derived using RMT tools.

In the absence of detailed balance, dynamical functionals can be used as an alternative, within a statistical mechanics approach to inference (for a systematic discussion see \cite{domany,engel}). 
The main result here is a generalization of \eqref{algcorr} to any degree of symmetry, which therefore provides important insights into the strength of non-equilibrium effects on the inference error.
We recall that the aim is to characterize a posterior path distribution, $P(\bm{X}^{\rm b}|\bm{X}^{\rm s})$, known to be Gaussian. 
The likelihood of the observed trajectory
$P(\bm{X}^{\rm s})$ can be seen as a ``partition function" $Z$ that is obtained by summing $P(\bm{X}^{\rm b}, \bm{X}^{\rm s})$ over all possible hidden paths 
$\bm{X}^{\rm b}$. From $Z$, one can define a free energy (density)
to study macroscopic quantities such as mean and covariance of $P(\bm{X}^{\rm b}|\bm{X}^{\rm s})$. If the interactions are chosen randomly, 
they act as quenched disorder and the physically relevant quantity is the 
quenched average of the free energy,
\be
f= -\text{lim}_{N \to \infty} N^{-1} \langle \ln Z({\bm{J},\bm{K}^{\rm sb}}) \rangle_{\bm{J},\bm{K}^{\rm sb}},
\ee
where we have abbreviated $N^{\rm b} \equiv N$. The free energy $-N^{-1}\ln Z$ is self-averaging, i.e.\ its fluctuations around $f$ for different realizations of the disorder vanish for $N\to\infty$. 
The same is true for the order parameters that arise in the calculation, which include the posterior variance, i.e.\ inference error.

Dynamical functionals appear in the above approach once we write the joint path probability $P(\bm{X}^{\rm b},\bm{X}^{\rm s})$ defined by the dynamics
\eqref{eq:lineq1} and \eqref{eq:lineq2} in Onsager-Machlup form as proportional to 
\begin{eqnarray}
\label{Onsager-Machlup}
&&P(\bm{X}^{\rm b},\bm{X}^{\rm s})\propto \\
&&\exp\left[-\frac{1}{2\sigma_{\rm b}^2}\int_0^T\big|\big|\partial_t\bm{x}^{\rm b}-\kbs\bm{x}^{\rm s}(t)-\jm^{\rm bb}\bm{x}^{\rm b}(t)\big|\big|^2dt\right]\notag\\
&&\cdot\exp\left[-\frac{1}{2\sigma_{\rm s}^2}\int_0^T\big|\big|\partial_t \bm{x}^{\rm s}-\jm^{\rm ss}\bm{x}^{\rm s}(t)-\ksb\bm{x}^{\rm b}(t)\big|\big|^2dt\right],\notag
\end{eqnarray}
with $\jm^{\rm bb}= -\lambda\mathbb{1} + \bm{J}$.
From the Gaussian form of this, the second order statistics of the posterior $P(\bm{X}^{\rm b}|\bm{X}^{\rm s})$ are independent of the value of the observed $\bm{X}^{\rm s}$.
Hence to obtain the posterior variance it is sufficient to consider {\em zero observations}, i.e.\ $x_a(t) =0$ for all $a$ and $t$. All $\bm{x}^{\rm b}$ are then effectively 
deviations $\delta\bm{x}^{\rm b}$ from the posterior mean, though we will not write the $\delta$ explicitly to save space. The only remaining
contribution from observations in \eqref{Onsager-Machlup} is in the couplings $K_{aj}$ and the relevant partition function becomes
\begin{equation}
Z = \left\langle\exp\left[-\frac{1}{2\sigma_{\rm s}^2} \sum_{a=1}^{N^{\rm s}} 
\int_0^T \left(\sum_{j=1}^N K_{aj} x_j(t)\right)^2 dt
\right]\right\rangle_{\bm{x}},
\end{equation}
where $\bm{x}\equiv \bm{x}^{\rm b}=\{x_i\}_{i=1}^N$. The average is the marginalization over the hidden dynamics with the weight given by the second term in \eqref{Onsager-Machlup}. 
This weight corresponds to the dynamics of the isolated hidden network, viz.
\begin{equation}
\partial_t x_i(t) = - \lambda x_i(t) + \sum_j J_{ij} x_j(t) + \xi_i(t),
\end{equation}
with white noise $\langle \xi_i(t) \xi_j(t')\rangle = \sigma^2_{\rm b} \delta_{ij}\delta(t - t')$ as before.

\subsection{Asymmetric hidden-hidden couplings}
\label{sec:asydyn}
\subsubsection{Annealed average}
The average of $\ln Z$ over the quenched couplings $\bm{J}$ and $\bm{K}^{\rm sb}$ would conventionally be performed by the replica method.
However, for fully connected systems with quadratic interaction terms such as the one here, similar calculations \cite{edjon,oppersolvable} indicate that the annealed calculation, which replaces $\langle \ln Z\rangle$ by $\ln \langle Z\rangle$, will give the exact result. We therefore calculate
\be
f= -\text{lim}_{N \to \infty}N^{-1}\ln \langle  Z(\bm{J},\bm{K}^{\rm sb}) \rangle_{\bm{J},\bm{K}^{\rm sb}}.
\ee

We shall again assume $\bm{J}$ and $\bm{K}^{\rm sb}$ to have Gaussian-distributed elements with zero mean, but now consider the case where $\bm{J}$ is {\em asymmetric}, i.e.\ $\langle J_{ij}J_{ji} \rangle=0$, thus breaking detailed balance. (We comment on the case of general symmetry of $\bm{J}$ below.) For the calculation we introduce
\begin{eqnarray}
\chi_i(t) &=& \sum_{j=1}^N J_{ij} x_j(t) + \xi_i(t),\\
\phi_a(t) &=& \sum_{j=1}^N K_{aj} x_j(t).
\end{eqnarray}
With regards to the quenched disorder average these are two Gaussian fields, which become independent when conditioned on the $x_i$.
Defining as before amplitudes $j$ and $k$ so that $\langle J_{ij}^2\rangle = j^2/N$ and $\langle K_{aj}^2\rangle = k^2/N$, we have
\begin{eqnarray}
\langle\chi_i(t) \chi_i(t')\rangle_{\bm{J}}
&=&\sigma^2_{\rm b} \delta(t - t') + j^2 C(t,t'),\label{covchi}\\
\langle\phi_a(t) \phi_b(t') \rangle_{\bm{J}} &=&  k^2 C(t,t') \delta_{ab},\label{covphi}
\end{eqnarray}
where we have introduced the order parameter
\begin{equation}
C(t,t')  \doteq \frac{1}{N} \sum_{j=1}^{N} x_j(t) x_j(t').
\end{equation}
Hence, we will calculate 
\begin{equation}
Z _{\rm ann} = \left\langle\exp\left[\frac{1}{2\sigma_{\rm s}^2} \sum_{a =1}^{N^{\rm s}} \int_0^T \phi^2_a(t) dt\right]\right\rangle_{\bm{\phi}, \bm{x}} ,
\end{equation}
where now the process has an effective prior dynamics given by 
\begin{equation}
\partial_t x_i(t)  = - \lambda x_i (t) +  \chi_i(t).
\label{eff_prior}
\end{equation}
Here $\bm{\phi}=\{\phi_a \}_{a=1}^{N^{\rm s}}$ and $\bm{\chi}=\{\chi_i\}_{i=1}^N$ are still coupled to $\bm{x}$
because of the covariances $C(t,t')$.

\subsubsection{Decoupling the degrees of freedom}
To decouple the degrees of freedom we constrain the value of the order parameter function $C(t, t')$. Formally this means writing $Z_{\rm ann}$ as an integral of $\exp(N\Xi[C])$ over
all possible values of $C(t,t')$, where
\begin{widetext}
\begin{eqnarray}
\label{G}
\Xi[C] = \frac{1}{N}\ln 
\left\langle\exp\left\{-\frac{1}{2\sigma_{\rm s}^2}  \sum_{a =1}^{N^{\rm s}} \int_0^T \phi^2_a(t) dt\right\} 
\prod_{t,t'} \delta\left(N C(t,t') - \sum_{i=1}^N x_i(t) x_i(t')\right) \right\rangle_{\bm{\phi}, \bm{x}}\equiv \Xi_1[C] + \Xi_2[C]
\end{eqnarray}
\end{widetext}
with 
\be
\label{G1}
\Xi_1[C] = \frac{1}{N} \ln \left\langle\prod_{t,t'} \delta\left(N C(t,t') - \sum_{i=1}^N x_i(t) x_i(t')\right) \right\rangle_{\bm{x}},
\ee
\be
\label{G2}
\Xi_2[C]=\frac{N^{\rm s}}{N} \ln \left\langle\exp\left\{-\frac{1}{2\sigma_{\rm s}^2}  \int_0^T \phi^2(t) dt\right\}\right\rangle_{\phi} .
\ee
In equation \eqref{G2} the decoupling has allowed us to drop the index $a$ and consider a representative $\phi$.

The first equation \eqref{G1} is dealt with by introducing an order parameter to $C(t,t')$. This means that for $N\to\infty$, we 
replace the ``hard" $\delta$ constraints by an extra Gaussian term yielding a new effective measure over independent $x_i(t)$, which is adjusted such that
$\langle x_i(t) x_i(t')\rangle_e = C(t,t')$ (here $e$ denotes the effective ``posterior" average). Equivalently one can write $\delta$-function constraints in Fourier representation and evaluate $\exp(N\Xi[C])$ using a saddle point method. Either way one has
\begin{widetext}
\begin{eqnarray}
\label{g1new}
\Xi_1= \frac{1}{2}\int_0^T dt  \int_0^T dt' \: \B(t,t') C(t,t')  + \ln \left\langle
\exp\left\{- \frac{1}{2}\int_0^T dt \int_0^T dt' \:  \B(t,t') x(t) x(t') \right\}\right\rangle_x.
\end{eqnarray}
\end{widetext}
This path integral is now also for a single representative coordinate $x$. Extremization over $\B(t,t')$ is understood in \eqref{g1new}, and similarly one needs to extremize over $C(t,t')$ in evaluating the resulting $Z_{\rm ann}$.

\subsubsection{Evaluating the order parameters}

As before we focus on the steady state of the system for $t\to\infty$.
The order parameters then depend on time differences only and the path integrals can be evaluated using Fourier or Laplace modes $\tilde{x}(z)$. These decouple into independent Gaussians
and we get from \eqref{covchi}, \eqref{covphi} and (\ref{eff_prior}) that 
\be
\label{eqc0}
\tilde{C}_0(z) \doteq \left\langle|\tilde{x}(z)|^2 \right\rangle_{\tilde{x}} = \frac{j^2 \tilde{C}(z) + \sigma_{\rm b}^2}{-z^2 + \lambda^2},
\ee
\be
\left\langle| \tilde{\phi}(z)|^2 \right\rangle_{\tilde{\phi}} = k^2 \tilde{C}(z).
\ee
$\tilde{C}_0(z)$ is the covariance of the prior effective dynamics while $\tilde{C}(z)$ relates to the posterior dynamics that includes the conditioning on observations.
Carrying out the prior average, the second term in \eqref{g1new} becomes 
\begin{eqnarray}
\label{secondg1}
&\ln& \left\langle \exp\left\{- \frac{1}{2}\int_0^T dt \int_0^T dt' \: \B(t,t') x(t) x(t') \right\}\right\rangle_{x}\notag\\
&=&-\frac{1}{2}\int dz \ln\bigg(1 + \tilde{C}_0(z)\tilde{\B}(z)\bigg).
\end{eqnarray}
In a similar way, we have for $\Xi_2$, from \eqref{G2} 
\begin{eqnarray}
\label{newg2}
\Xi_2&=&\left\langle\exp\left\{-\frac{1}{2\sigma_{\rm s}^2}  \int_0^T \phi^2(t) dt\right\} \right\rangle_\phi \notag\\
&=&- \frac{1}{2}\int dz \ln\left(1 + \frac{k^2}{\sigma_{\rm s}^2} \tilde{C}(z)\right).
\end{eqnarray}
Hence, finally, by substituting \eqref{secondg1} into \eqref{g1new} and from \eqref{newg2} we get
\begin{widetext}
\be
\Xi = \frac{1}{2} \int dz \left[\tilde{\B}(z)\tilde{C}(z) - \ln\bigg(1 + \tilde{C}_0(z)\tilde{\B}(z)\bigg)\right]-\frac{\alpha}{2}\int dz \ln\left(1 +
\frac{k^2}{\sigma_{\rm s}^2} \tilde{C}(z)\right),
\ee
\end{widetext}
where $\alpha=N^{\rm s}/N$ as before.
The order parameter equations $\partial \Xi/ \partial \tilde{C}(z)=0$ and $\partial \Xi/ \partial\tilde{\B}(z)=0$ result as
\be
\label{finalpos}
\tilde{\B}(z) =\frac{\alpha k^2}{\sigma_{\rm s}^2 + k^2 \tilde{C}(z)}
+ \frac{\tilde{\B}(z)}{1 + \tilde{C}_0(z)\tilde{\B}(z)}\frac{j^2}{-z^2 + \lambda^2},
\ee
\be
\label{finalpos1}
\frac{\tilde{C}(z)}{\tilde{C}_0(z)} + \tilde{\B}(z)\tilde{C}(z) = 1.
\ee
Combining these and using \eqref{eqc0} gives a closed algebraic equation for $\tilde{C}(z)$
\clearpage
\begin{eqnarray}
\label{algcorr0}
&&z^2=\\
&&\left[-\frac{\sigma_{\rm b}^2}{\tilde{C}} +\frac{\alpha \frac{k^2 \sigma_{\rm b}^2}{\sigma_{\rm s}^2}}
{1+ \frac{k^2}{\sigma_{\rm s}^2}\tilde{C}}\right]\bigg(1+\frac{j^2}{\sigma_{\rm b}^2}\tilde{C}\bigg)^2
+j^2 \bigg(1+\frac{j^2}{\sigma_{\rm b}^2}\tilde{C}\bigg)+ \lambda^2\notag
\end{eqnarray}
with the abbreviation $\tilde{C}(z)=\tilde{C}$. This is the analog of \eqref{algcorr} for the non-equilibrium case of 
asymmetric couplings $\bm{J}$, and our final result for this section.\\

\subsection{Generalization to arbitrary interaction symmetry}
\label{sec:gensym}
The above approach based on dynamical functionals can be extended to the case of hidden-hidden interactions of arbitrary degree of symmetry, defined by $\langle J_{ij}J_{ji} \rangle = \eta j^2/N$. Asymmetric couplings (section \ref{sec:asydyn}) correspond to $\eta=0$ while
$\eta=1$ gives symmetric $\bm{J}$ (section \ref{sec:symmetric_couplings}).
We do not detail the calculations for the case of general $\eta$ here. The main change is that the nonzero correlation $\langle J_{ij}J_{ji} \rangle$ causes
the effective prior dynamics to contain a response term where each $x_i(t)$ reacts to its values $x_i(t')$ in the past (see e.g.\ \cite{sompolinsky1}).

The final result is again a closed algebraic equation for $\tilde{C}(z)$
\begin{widetext}
\be
\label{algcorr_gen}
z^2= \left[-\frac{\sigma_{\rm b}^2}{\tilde{C}} +\frac{\alpha \frac{k^2 \sigma_{\rm b}^2}{\sigma_{\rm s}^2}}{1+ \frac{k^2}{\sigma_{\rm s}^2}\tilde{C}}
+\frac{j^2}{1+\frac{j^2}{\sigma_{\rm b}^2}\tilde{C}}+ 
\frac{\lambda^2}{\bigg(1+(1+\eta)\frac{j^2}{\sigma_{\rm b}^2}\tilde{C}\bigg)^2}\right]
\bigg(1+(1-\eta)\frac{j^2}{\sigma_{\rm b}^2}\tilde{C}\bigg)^2.
\ee
\end{widetext}
For $\eta=1$ and $\eta=0$ this leads back to \eqref{algcorr} and \eqref{algcorr0}, respectively, as it should.

The result \eqref{algcorr_gen} characterizes the average case posterior variance -- and hence inference error -- for our partially observed network dynamics. 
Remarkably, it does so across an entire range of non-equilibrium settings parameterized by $\eta$. Equation \eqref{algcorr_gen} is derived within the annealed approximation 
but as discussed above this should be exact here so that our result acts as a baseline for the assessment of other approximations. One such approximation, 
the Extended Plefka Expansion~\cite{plefkaobs,plefkaobs2}, can be shown to give exactly \eqref{algcorr_gen}, demonstrating 
that this approximate scheme is also exact (in the large system limit studied here).

The dependence on various parameters, especially the level of symmetry $\eta$, of 
inference errors and posterior relaxation times as they result from \eqref{algcorr_gen}
 is sufficiently rich that we devote a separate paper to it~\cite{plefkaobs2}.
It turns out that the behavior can be organized around critical regions in the parameter space of $\alpha$, $\gamma$ and $p$. There are two such regions. Generalizing from section \ref{sec:post_symm}, these are defined by $p\to 0$ for $0\leq \alpha \leq 1$ for the first region, and for the second $\alpha\to 0$ and
$\gamma \to \gamma_{\rm c}=1/(1+\eta)$. One key finding is that across the entire range of eta from 0 to just below 1, i.e. the regime where  interaction symmetry is broken, there are no qualitative changes in behavior. 
On the other hand, interesting crossovers then occur in the vicinity of $\eta=1$, i.e.\ as interaction symmetry is approached. We refer the interested reader to 
\cite{plefkaobs2} for further details.

\section{Discussion and Conclusions}
\label{Discussion}

We have considered in this paper linear stochastic dynamics in a large network of continuous degrees of freedom, where given a time trajectory of the nodes in some observable part of the network the task is to infer the trajectory of the hidden nodes. 
By varying interaction symmetry we were able to study both equilibrium and non-equilibrium settings, thus creating a paradigmatic example of inference from temporal data.
Given the increasing availability of large scale temporal data sets such problems are becoming prevalent in e.g.\ biology, 
where interpretation of data and prediction are highly challenging when observations only partially characterize a system.

Our main goal was to explore the average case inference error. To ensure analytical tractability we focused on stationary dynamics on large networks. More precisely it is the variance of hidden state estimates that becomes stationary in time; mean predictions for the hidden states have to depend on time in our dynamical context.
The large network assumption is realistic in many situations, e.g.\ for metabolic or neural networks that
can be composed of thousands of interacting elements (chemical species,  neurons etc).

We deployed two different methods of analysis. For the first,
the starting point (section \ref{Stationary_Posterior_Variance}) is a Lyapunov-type equation for the 
posterior variance matrix $\bm{C}$, where an effective drift matrix $\kbbs$ captures the effect of the observations. 
In section \ref{sec:RMT} we derived average case performance results by appeal to RMT. This is possible because the Lyapunov equation can be solved in the case of self-interacting hidden variables (section 
\ref{sec:selfinteraction}) or more generally, symmetric hidden-hidden couplings (section \ref{sec:symmetric_couplings}), corresponding to equilibrium dynamics. With suitable assumptions of couplings being Gaussian and long-range, and taking the thermodynamic limit of large networks, we then used free probability methods to derive the 
Green's functions and then the spectra of $\bm{C}$ and $\kbbs$, which are closely linked.

For the opposite case of {\em asymmetric} hidden-hidden couplings, where the dynamics is non-equilibrium, we presented in section \ref{sec:asydyn} a calculation based on dynamical functionals. This leads
to an algebraic equation for the stationary posterior variance (in Laplace space). We sketched how the approach can be extended to the analysis of 
non-equilibrium stationary regimes arising 
from couplings of {\em generic} symmetry (section \ref{sec:gensym}).

We focused on the inference error as an average macroscopic quantity. For large networks this is independent of the specific realization of the microscopic (Gaussian) interactions, but does depend on 
structural parameters such as overall interaction strengths 
as well as $\alpha$, the ratio between the number of hidden and observed nodes. 
Predictions on such structural dependences of macroscopic properties should be testable in practice
and may give information on microscopic features such as the degree of interaction symmetry. 
The emerging picture, consisting of algebraic expressions that link inference
errors and parameters, suggests possible connections to experiment design, as we discuss further in \cite{plefkaobs2}. 
There we quantify these dependences in terms of scaling laws; of particular importance is the dependence
on $\alpha$, as it tells us how many observed nodes are needed to attain a specified precision for the hidden node inference. 

The RMT approach to our problem has the benefit that it gives information on spectral densities - our main focus here - 
including the spectrum of relaxation times in the posterior dynamics. 
This then allowed us to compare different definitions of a characteristic posterior relaxation time, such as slowest mode and average time (section \ref{sec:justself}).
The spectral shapes proved revealing: when there are few observations (small $\alpha$), the spectrum can be split into two parts corresponding to constrained and unconstrained directions 
(section \ref{sec:symmetric_couplings}), but this distinction is then lost as hidden nodes interact more strongly.

One open question for the inference setting we have considered is to answer the question of the spectral density of relaxation 
times and its support in the {\em non-equilibrium} case $\eta<1$. 
For example, does our result  \eqref{algcorr_gen} for generic $\eta$ still 
have a free probability interpretation? Generalizing the derivation of the equilibrium ($\eta=1$) result \eqref{algcorr} to $\eta<1$ appears non-trivial.
One might consider assuming that the equilibrium relation
$\tilde{C}(z)= -\sigma_{\rm b}^2 \tilde{G}(z^2)$ continues to hold and analyze the spectrum corresponding to the Green's function $\tilde{G}(z)$.

There are a number of avenues for further work, as the setting we have begun to study is still rather
new in the statistical physics community \cite{romanobattistin,inference1,inference2,roudi1}. 
An obvious extension would be to sparse networks, where for static analyses statistical mechanics has been successfully deployed \cite{Reimer,Rogers}.
The sparse case would be worth developing because of its relevance to applications such as gene expression networks \cite{braunstein}.
As a starting point one could investigate progressive 
degrees of dilution. Consider for example an average degree of connectivity $c$, which corresponds
to the $J_{ij}$ being drawn as Gaussian random variables with probability $c/N$, and 
zero with probability $1-c/N$; 
one would set then the amplitude of
the nonzero $J_{ij}$ such that $\langle J_{ij}^2\rangle =j^2/c$ in order to obtain
a sensible thermodynamic limit. In this paper, we have effectively considered $c = N$, but from previous studies \cite{Erd+12,Erd+13} it is clear that
one can take $c \ll N$ (in fact as low as $c\sim \ln N$) 
without changing the results derived in this paper. This already goes a long way towards making our work applicable to real networks. The strong dilution regime, where $c=O(1)$, would require
a separate analysis that goes beyond the scope of the present paper. Cavity and population dynamics methods developed for sparse network spectra 
(e.g.~\cite{Reimer,Rogers}) would probably need to be deployed there. 

A second important consideration for applications to real networks is their finite size $N$. We have begun to investigate the resulting finite size effects numerically. 
Encouragingly, we find \cite{plefkaobs} that even for moderate network sizes ($N\approx 100$) there is good agreement between numerically exact calculations of the inference error on the one hand and our 
large-$N$ theory on the other.

Variants of the dynamics could also be considered, for example, by adding non-linearities that can be treated perturbatively. 
One could also extend to measurements of the trajectory of the observable nodes that would be available at a regular or irregular grid of time 
points only rather than along the entire time interval considered; or to measurements which are noisy rather than just incomplete as in our 
case \cite{opper_gaussian,oppersangui2}.

Finally, we have concentrated on the {\em forward} problem of predicting hidden states given known interactions. 
This is relevant also for inverse problems such as learning the couplings from dynamical data, where typically a forward problem has to be solved at every
iteration (e.g.\ in Expectation Propagation \cite{opperwinther}). Learning which couplings are non-zero 
is effectively a network reconstruction problem, with potential applications 
to signaling pathways and gene expression data. In either case, modelling data as explicitly dynamical rather than as uncorrelated snapshots is expected to lead to performance improvements in inference and learning.
Such algorithmic advances have already been achieved by adapting equilibrium statistical physics tools \cite{perturbations, braunstein} to learning of regulatory networks from steady state data.

\section*{Acknowledgements}
This work was supported by the Marie Curie Training Network NETADIS (FP7, grant 290038). We are grateful to Pierpaolo Vivo, Ludovica Bachschmid-Romano
and Reimer K\"{u}hn for helpful discussions.

\appendix
\section{Kalman filter and smoother}
\label{appendix:kalman}
In this appendix we derive the results \eqref{eq:post}-\eqref{postdecay} in the main text, using a reduction of our inference problem to a linear 
Gaussian state space model, to which standard Kalman filter techniques~\cite{bishop} can then be applied.

Let us consider a time discretized version of our dynamics \eqref{eq:lineq1} and \eqref{eq:lineq2}, with elementary time step $\Delta$,
\begin{eqnarray}
\label{eq:lineq1dis}
&&\bm{x}^{\rm b}(t)-\bm{x}^{\rm b}(t-\Delta)=\\
&&\Delta \kbs\bm{x}^{\rm s}(t-\Delta)+\Delta \jm^{\rm bb}\bm{x}^{\rm b}(t-\Delta)+\Delta \bar{\bm{\xi}}^{\rm b}(t-\Delta),\nonumber
\end{eqnarray}
\begin{eqnarray}
\label{eq:lineq2dis}
&&\bm{x}^{\rm s}(t)-\bm{x}^{\rm s}(t-\Delta)=\\
&&\Delta \jm^{\rm ss}\bm{x}^{\rm s}(t-\Delta)+\Delta \ksb\bm{x}^{\rm b}(t-\Delta)+\Delta \bar{\bm{\xi}}^{\rm s}(t-\Delta)\nonumber,
\end{eqnarray}
where the white noises $\bar{\bm{\xi}}^{\rm s}$ and $\bar{\bm{\xi}}^{\rm b}$ are averages of the continuous time noise over the time interval $\Delta$ with covariance 
\begin{equation}
 \langle \bar{\bm{\xi}}^{\rm s}(t)\bar{\bm{\xi}}^{\rm s\,\it{T}}(t')\rangle=\Delta^{-1}\bm{\Sigma}^{\rm ss}\delta_{tt'}
\end{equation}
and similarly for $\bar{\bm{\xi}}^{\rm b}$.
The above dynamics is Markovian, with transition probabilities
\small
\begin{eqnarray}
&&P(\bm{x}^{\rm b}(t)|\bm{x}^{\rm b}(t-\Delta),\bm{x}^{\rm s}(t-\Delta))=\\
&&\mathcal{N}(\bm{x}^{\rm b}(t)|(\mathbb{1}+\Delta \jm^{\rm bb})\bm{x}^{\rm b}(t-\Delta)+
\Delta \jm^{\rm sb}\bm{x}^{\rm s}(t-\Delta),\Delta \bm{\Sigma}^{\rm bb})\label{trGaufull},\nonumber\\
&&P(\bm{x}^{\rm s}(t+\Delta)|\bm{x}^{\rm b}(t),\bm{x}^{\rm s}(t))=\\
&&\mathcal{N}(\bm{x}^{\rm s}(t+\Delta)|(\mathbb{1}+\Delta\bm{K}^{\rm ss})\bm{x}^{\rm s}(t)+\Delta\ksb\bm{x}^{\rm b}(t),\Delta \bm{\Sigma}^{\rm ss})
\label{emGaufull}\nonumber
\end{eqnarray}
\normalsize
and we are interested in the posterior probability $P({\bm{X}}^{\rm b}|{\bm{X}}^{\rm s})$ of a time trajectory 
${\bm{X}}^{\rm b}$ of hidden variables given a trajectory ${\bm{X}}^{\rm s}$ of observed variables. 

To bring this inference problem into a standard form, we exploit the fact that the joint distribution 
$P({\bm{X}}^{\rm b},{\bm{X}}^{\rm s})$ is Gaussian, and hence so is the posterior $P({\bm{X}}^{\rm b}|{\bm{X}}^{\rm s})$. 
From general properties of Gaussian conditioning, the second order statistics of the posterior are then {\em independent} of the specific observed 
trajectory ${\bm{X}}^{\rm s}$. 
We can therefore choose the most convenient ${\bm{X}}^{\rm s}$ to find the second order statistics, which is the identically zero trajectory. 
The second order statistics we find then determine the inference error, which is the trace of the covariance matrix of ${\bm x}^{\rm b}(t)$.
 \begin{figure}
 \begin{center}
\includegraphics[width=0.48\textwidth]{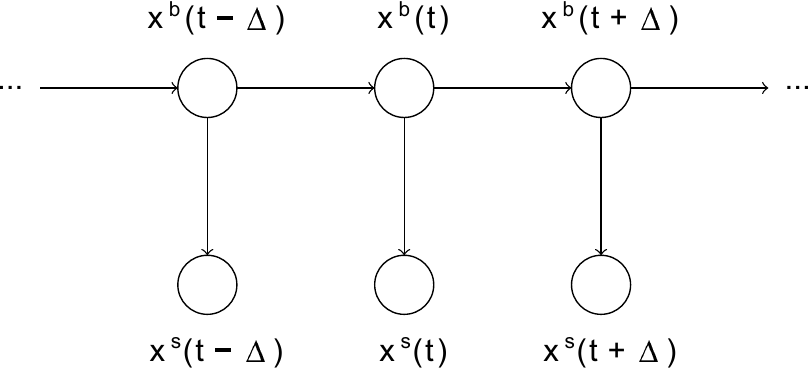}
  \caption{\footnotesize Illustration of a linear-Gaussian state space model.}
\label{fig:lGsm}
\end{center}
\end{figure}
For zero observations, the transition probabilities \eqref{trGaufull}, \eqref{emGaufull} simplify to
\begin{eqnarray}
&&{}P(\bm{x}^{\rm b}(t)|\bm{x}^{\rm b}(t-\Delta))=\label{trGau}\\
&&{}\mathcal{N}(\bm{x}^{\rm b}(t)|(\mathbb{1}+\Delta \jm^{\rm bb})\bm{x}^{\rm b}(t-\Delta),\Delta \bm{\Sigma}^{\rm bb}),\nonumber\\
&&{}P(\bm{x}^{\rm s}(t+\Delta)=0|\bm{x}^{\rm b}(t))= \label{emGau}\\
&&\mathcal{N}(\bm{x}^{\rm s}(t+\Delta)=0|\Delta\ksb\bm{x}^{\rm b}(t),\Delta \bm{\Sigma}^{\rm ss}).\nonumber
\end{eqnarray}
These now have the conventional form of a linear-Gaussian state space model
\cite{bishop}, where \eqref{trGau} specifies the dynamics of the hidden state ${\bm{x}}^{\rm b}$ while \eqref{emGau} defines the 
``emission probability'' at time $t$, with $\bm{x}^{\rm s}(t+\Delta)$ taking the role of the emitted signal or observation. 
To conform with standard notation, we will shift the time index on $\bm{x}^{\rm s}(t+\Delta)$ to $\bm{x}^{\rm s}(t)$ for the rest of this discussion; see figure \ref{fig:lGsm}.
Note that while
we are dealing with real-valued states and emissions here, the probabilistic ``graphical model''~\cite{bishop} of figure
\ref{fig:lGsm} could also capture cases, e.g.\ {\em Hidden Markov Models} (HMMs) where the hidden states are discrete.

The chain structure of figure~\ref{fig:lGsm} means that 
posterior probabilities can be computed efficiently by message passing methods,  denoted {\em Forward-Backward} algorithm in 
the context of HMMs \cite{rabiner}
and {\em Kalman Filter} \cite{kalmanor}
\footnote{Rigorously only the recursive computation of forward messages should be referred to as Kalman filter \cite{kalmanor}, while equations
of backward messages are known as Kalman smoothers.} here.

The forward propagation computes forward messages $\hat\alpha_t$ that absorb the effect of previous observations (the past), while the backward propagation accounts for observations from the future. Formally the messages can be defined as
\be
\hat{\alpha}(\bm{x}^{\rm b}(t))=P(\bm{x}^{\rm b}(t)|\bm{x}^{\rm s}(\Delta),...,\bm{x}^{\rm s}(t))=\hat{\alpha}_t,
\ee
\begin{eqnarray}
\hat{\beta}(\bm{x}^{\rm b}(t))&=& \frac{P(\bm{x}^{\rm s}(t+\Delta),...,\bm{x}^{\rm s}(T)|\bm{x}^{\rm b}(t))}{P(\bm{x}^{\rm s}
(t+\Delta),...,\bm{x}^{\rm s}(T)|\bm{x}^{\rm s}(\Delta),...,\bm{x}^{\rm s}(t))}\nonumber\\
&=&\hat{\beta}_t.
\end{eqnarray}
Once $\hat{\alpha}_t$ and $\hat{\beta}_t$ have been computed, the desired 
posterior probability is simply
\be
\label{gammat}
\gamma_t=\hat{\alpha}_t\hat{\beta}_t= \frac{P(\bm{x}^{\rm b}(t),\bm{X}^{\rm s})}{P(\bm{X}^{\rm s})}=P(\bm{x}^{\rm b}(t)|\bm{X}^{\rm s}).
\ee

The forward propagation for continuous variables reads
\begin{eqnarray}
\label{forprop}
\hat{\alpha}_t&\propto & P(\bm{x}^{\rm s}(t)|\bm{x}^{\rm b}(t)) \cdot\\
&&\int d\bm{x}^{\rm b}(t-\Delta) P(\bm{x}^{\rm b}(t)|\bm{x}^{\rm b}(t-\Delta))\hat{\alpha}_{t-\Delta}.\nonumber
\end{eqnarray}
In our case, all distributions involved are Gaussian and we denote in particular
\be
\label{alphaGau}
\hat{\alpha}_t= \mathcal{N}(\bm{x}^{\rm b}(t)|0,\cf(t)).
\ee
$\cf(t) = \langle \bm{x}^{\rm b}(t)\bm{x}^{\rm b}(t)^T\rangle$ is the equal time forward (or ``filtered'') posterior covariance.
By substituting \eqref{trGau}, \eqref{emGau} and \eqref{alphaGau} into
\eqref{forprop} and identifying the quadratic terms in $\bm{x}^{\rm b}(t)$ in the exponents one obtains the recursive Kalman filter expression 
for $\cfsi(t)$
\begin{eqnarray}
\label{eq:cfprop}
\cfsi(t)&=&\big[(\mathbb{1}+\Delta\jm^{\rm bb})\,\cf(t-\Delta) \big(\mathbb{1}+\Delta\jm^{\rm bb}\big)^T\nonumber\\
&+&\Delta\bm{\Sigma}^{\rm bb}\big]^{-1}+\Delta\bm{W},
\end{eqnarray}
where $\bm{W}=\bm{K}^{\rm sb\,\it{T}}(\bm{\Sigma}^{\rm ss})^{-1}\ksb$ is the \emph{feedback} matrix.
Equation \eqref{eq:cfprop} is a discrete time Riccati (i.e.\ second order matrix) recursion. We are interested in the continuous time limit $\Delta \rightarrow 0$, where it becomes
\begin{eqnarray}
\label{eq:cfi}
 &&\frac{d}{dt}\cfsi(t)=\\
 &&\cfsi(t)\bm{\Sigma}^{\rm bb}\cfsi(t)+\cfsi(t)\jm^{\rm bb}+\jm^{\rm bb\,\it{T}}\cfsi(t)+\bm{W}.\nonumber
\end{eqnarray}

The backward propagation incorporates in the algorithm the observations from all later time steps
\begin{eqnarray}
\label{backprop}
\hat{\beta}_t\propto \int d\bm{x}^{\rm b}(t+\Delta)&&\hat{\beta}_{t+\Delta}P(\bm{x}^{\rm s}(t+\Delta)|\bm{x}^{\rm b}(t+\Delta))\nonumber\\
&&\cdot P(\bm{x}^{\rm b}(t+\Delta)|\bm{x}^{\rm b}(t))
\end{eqnarray}
and we set
\be
\label{betaGau}
\hat{\beta}_t\propto \mathcal{N}(\bm{x}^{\rm b}(t)|0,\cb(t))
\ee
with $\cb(t)= \langle \bm{x}^{\rm b}(t)\bm{x}^{\rm b}(t)^T\rangle$ defined as the equal time posterior variance in the 
backward propagation. Inserting \eqref{betaGau} into \eqref{backprop} one finds the backward recursion for $\cbi(t)$ 
\begin{eqnarray}
\cbi(t)&=&\big(\mathbb{1}+\Delta\jm^{\rm bb}\big)^T(\Delta \bm{\Sigma}^{\rm bb})^{-1}\cdot\\
&&\big[\mathbb{1} -\big(\mathbb{1}+
\Delta\bm{\Sigma}^{\rm bb} \cbi(t+\Delta)+\Delta^2 \bm{\Sigma}^{\rm bb}\bm{W}\big)^{-1}\big]\nonumber\\
&&\cdot(\mathbb{1}+\Delta\jm^{\rm bb}).\nonumber
\end{eqnarray}
Taking $\Delta\to0$, which requires keeping all terms up to $O(\Delta)$ on the r.h.s., gives the continuous time limit
\begin{eqnarray}
\label{eq:cbi}
&&\frac{d}{dt}\cbi(t)=\\
&&-\jm^{\rm bb\,\it{T}} \cbi(t) - \cbi(t) \jm^{\rm bb} - \bm{W} +\cbi(t)\bm{\Sigma}^{\rm bb}\cbi(t).\nonumber
\end{eqnarray}
The changes of sign compared to \eqref{eq:cfi} come from the backward direction.

Finally the posterior $\gamma_t$ also has a Gaussian form,
\be
\label{gammaGau}
\gamma_t=\mathcal{N}(\bm{x}^{\rm b}(t)|0,\bm{C}^{\rm bb|s}(t)).
\ee
We drop the superscripts on $\bm{C}^{\rm bb|s}(t)$ as in the main text and write this overall (``smoothed'') covariance as $\bm{C}(t)$. From \eqref{gammat} one has $\bm{C}^{-1}(t)= \cfsi(t) +\cbi(t)$, so from the sum of \eqref{eq:cfi} and \eqref{eq:cbi}
\begin{eqnarray}
\label{eqc-1}
&&\frac{d}{dt}\bm{C}^{-1}(t) = \\
&&\bm{C}^{-1}(t)\bm{\Sigma}^{\rm bb}\bm{C}^{-1}(t)+ \bm{C}^{-1}(t)\jm^{\rm bb|s}+\jm^{\rm bb|s\,\it{T}} \bm{C}^{-1}(t), \nonumber
\end{eqnarray}
where we have set 
\be
\label{postdri}
\jm^{\rm bb|s}= \jm^{\rm bb}-\bm{\Sigma}^{\rm bb} \cbi
\ee
and we have taken $\cbi$ as the stationary limit of $\cbi(t)$.

To interpret $\jm^{\rm bb|s}$ one can look at $P(\bm{x}^{\rm b}(t+\Delta),\bm{x}^{\rm b}(t)|\bm{X}^{\rm s})$, given by the integrand of 
\eqref{backprop}. Conditioning on $\bm{x}^{\rm b}(t)$ and using \eqref{trGau}, \eqref{emGau} and \eqref{betaGau} one finds easily that
the 
mean of $\bm{x}^{\rm b}(t+\Delta)$ conditioned on $\bm{x}^{\rm b}(t)$ is
\begin{eqnarray}
\label{xpost}
\big(\mathbb{1} + \Delta \jm^{\rm bb|s}(t) + O(\Delta^2) \big)\bm{x}^{\rm b}(t).
\end{eqnarray} 
Hence $\jm^{\rm bb|s}(t)$ has the meaning
of a posterior drift, i.e. it determines the time evolution for the posterior dynamics.

Focusing on the stationary state now, we can drop all dependences on $t$. From \eqref{eqc-1}, the posterior covariance $\bm{C}$ then satisfies the Lyapunov equation \eqref{eq:post}
\begin{equation}
\label{eqLyap0}
 \jm^{\rm bb|s}\bm{C}+ \bm{C}\jm^{\rm bb|s\,\it{T}}+\bm{\Sigma}^{\rm bb}=0
\end{equation} 
with the stationary posterior drift $\kbbs$ given by
\be
\label{postdristat}
\jm^{\rm bb|s}= \jm^{\rm bb}-\bm{\Sigma}^{\rm bb} \cbi
\ee
and the stationary backward covariance satisfying, from \eqref{eq:cbi} 
\be
\cbi\bm{\Sigma}^{\rm bb}\cbi-\jm^{\rm bb\,\it{T}} \cbi - \cbi \jm^{\rm bb} =\bm{W}.
\ee
Apart from the relabelling of $\cbi$ as $\bm{A}$, we have therefore derived \eqref{eq:post}, \eqref{eq:postdrift} and \eqref{eq:riccati1} in the main text.
Note that $\cbi$ is symmetric by definition; it is also positive semi-definite. As it enters the effective drift with a minus sign, we see that the presence of observations drives the hidden dynamics back towards its mean (zero) more quickly.

To find the evolution of the two-time posterior variance $\bm{C}(t,t')$, we first look at the case $\bm{C}(t'+\Delta,t')$ of
adjacent time steps. Here \eqref{xpost} gives directly 
\be
\bm{C}(t'+\Delta,t') = \big(\mathbb{1} + \Delta \jm^{\rm bb|s}(t') + O(\Delta^2) 
\big) \bm{C}(t',t').
\ee
This easily generalizes to the correlations $\tau$ steps apart as
\be
\bm{C}(t'+\tau\Delta,t') = \big(\mathbb{1} + \Delta \jm^{\rm bb|s} + O(\Delta^2) \big)^{\tau} \bm{C},
\ee
where we have directly written the stationary version. 
Setting $t=t'+\tau\Delta$ and taking $\Delta\to 0$ then gives equation \eqref{postdecay} in the main text, i.e.\
\begin{equation}
\label{eq:exDe}
\bm{C}(t-t')= e^{\bm{K}^{\rm bb|s}(t-t')}\bm{C}.
\end{equation}

\section{Variational method}
\label{appendix:variational}
As is often the case, the fixed point of a recursion (such as the Forward-Backward algorithm) can also be retrieved 
variationally, i.e.\ as the solution of a constrained optimization problem. We show this connection in this appendix.

Let us start from $P(\bm{X}^{\rm b},\bm{X}^{\rm s})$, the joint probability of subnetwork and bulk trajectories 
obeying \eqref{eq:lineq1} and \eqref{eq:lineq2}, and denote
$Q(\bm{X}^{\rm b})$ a variational approximation to the posterior 
$P(\bm{X}^{\rm b}|\bm{X}^{\rm s})$ of the effective dynamics \eqref{eq:efflineq1}. 
As before if we are interested only in the posterior second order statistics, we can remove the means by 
assuming $\bm{x}^{\rm s}(t)=0$ $\forall t$ and can then drop the $\delta$ in \eqref{eq:efflineq1}.
One aim is to determine the effective drift $\kbbs$ by variational methods. Note that parameterizing $Q$ in terms of $\kbbs$ gives us enough flexibility to retrieve the {\em exact} posterior because of the Gaussian nature of our problem. 

We can write the joint trajectory probability and the variational posterior, directly in continuous time form, as 
\begin{eqnarray}
\label{eqPapp}
&&P(\bm{X}^{\rm b},\bm{X}^{\rm s})\propto\\
&& \exp\left[-\frac{1}{2}\int_0^Tdt\big(\bm{\xi}^{\rm b\,\it{T}}(t)\bm{\Sigma}^{\rm bb\,-1}\bm{\xi}^{\rm b}(t)+
\bm{\xi}^{\rm s\,\it{T}}(t)\bm{\Sigma}^{\rm ss\,-1}\bm{\xi}^{\rm s}(t)\big)\right]\nonumber
\end{eqnarray}
\be
\label{eqQapp}
Q(\bm{X}^{\rm b})\propto \exp\left[-\frac{1}{2}\int_0^Tdt\,\bm{\xi}^{\rm b\,\it{T}}(t)\bm{\Sigma}^{\rm bb\,-1}\bm{\xi}^{\rm b}(t)\right],
\ee
where the noises $\bm{\xi}^{\rm b}$ and $\bm{\xi}^{\rm s}$
should be expressed as a function of $\bm{x}^{\rm b}$ and $\bm{x}^{\rm s}$
using respectively equations \eqref{eq:lineq1} and \eqref{eq:lineq2} for $P(\bm{X}^{\rm b},\bm{X}^{\rm s})$ and
\eqref{eq:efflineq1} for $Q(\bm{X}^{\rm b})$.\\
We find $Q$ in the standard variational way by finding the stationary point of the Kullback-Leibler divergence \cite{kullback} between $P$ and $Q$ 
\be
\label{klapp}
\text{KL}(P||Q) = -\bigg\langle\log{\frac{Q}{P}}\bigg\rangle_Q= F,
\ee
which is analogous to a thermodynamic free energy. 
Inserting \eqref{eqPapp} and \eqref{eqQapp} and simplifying gives
\begin{widetext}
\be
F=\int_0^T dt \,\frac{1}{2}\left\langle\bm{x}^{\rm b\,\it{T}}(t)(\jm^{\rm bb}-\kbbs)^{T} \bm{\Sigma}^{\rm bb\,-1}(\jm^{\rm bb}-\kbbs)\bm{x}^{\rm b}(t) \right\rangle_Q
+ \int_0^T dt \, \frac{1}{2}\left\langle\bm{x}^{\rm b\,\it{T}}(t)\bm{W} \bm{x}^{\rm b}(t)\right\rangle_Q 
\ee
\end{widetext}
with $\bm{W}\doteq (\ksb)^T\bm{\Sigma}^{\rm ss\,-1}\ksb$ the feedback matrix as before. Here  we have performed an integration by parts 
and assumed that $\bm{x}^{\rm b}$ vanishes at the boundaries of the time domain. 

In the stationary limit, we can drop the time integrals, drop the resulting factor $T$
and use the definition $\bm{C} = \langle\bm{x}^{\rm b}\bm{x}^{\rm b\,\it{T}}\rangle_Q$ to write
\begin{eqnarray}
F &=& \frac{1}{2} \text{Tr}\left[(\jm^{\rm bb}-\kbbs)^{T}\bm{\Sigma}^{\rm bb\,-1} (\jm^{\rm bb}-\kbbs)\bm{C}\right] \nonumber\\
&+& \frac{1}{2}\text{Tr}\left[\bm{W} \bm{C} \right].
\end{eqnarray}
We now want to optimize over $\kbbs$, bearing in mind that 
the stationary posterior variance $\bm{C}$ is linked to the effective drift by the Lyapunov equation
\be
\label{postLyap}
\jm^{\rm bb|s}\bm{C}+ \bm{C}\jm^{\rm{bb|s}\,\it{T}}+\bm{\Sigma}^{\rm bb}=0
\ee
(see \eqref{eq:post} in the main text). Introducing a Lagrange multiplier matrix $\bm{A}/2$ to implement this constraint,
we optimize
\begin{eqnarray}
&&\mathcal{L}[\bm{C},\kbbs,\bm{A}]= \\
&&F+\frac{1}{2} \text{Tr}\left[\bm{A}^T(\jm^{\rm bb|s}\bm{C}+ \bm{C}\jm^{\rm{bb|s}\,\it{T}}+\bm{\Sigma}^{\rm bb})\right].\notag
\end{eqnarray}
Optimization w.r.t.\ $\kbbs$ gives
\be
\frac{\partial\mathcal{L}}{\partial \kbbs}= \bm{\Sigma}^{\rm bb\,-1}(\kbbs-\jm^{\rm bb})\bm{C}+\frac{1}{2}(\bm{A}+\bm{A}^T)\bm{C}=0,
\ee
from which one has the expression \eqref{eq:postdrift} for the posterior drift matrix
\be
\label{postdrift}
\kbbs = \jm^{\rm bb}-\frac{\bm{\Sigma}^{\rm bb}}{2}(\bm{A}+\bm{A}^T)= \jm^{\rm bb}- \bm{\Sigma}^{\rm bb} \bm{A}_{\rm s},
\ee 
where we have denoted the symmetric part of $\bm{A}$ by $\bm{A}_{\rm s}=\frac{1}{2}(\bm{A}+\bm{A}^T)$. We will then write
$\bm{A} = \bm{A}_{\rm s}+\bm{A}_{\rm a}$ with $\bm{A}_{\rm a} = \frac{1}{2}(\bm{A}-\bm{A}^T)$ the antisymmetric part.
The second optimization condition reads
\begin{eqnarray}
\label{secder}
\frac{\partial \mathcal{L}}{\partial \bm{C}}&=&
\frac{1}{2}(\kbbs-\jm^{\rm bb})^{T}\bm{\Sigma}^{\rm bb\,-1}(\kbbs-\jm^{\rm bb})\notag\\
&+& \frac{1}{2}\bm{W}+\frac{1}{2}(\bm{A}\kbbs+ \bm{K}^{\rm bb|s\,\it{T}}\bm{A})=0.
\end{eqnarray}
By substitution of \eqref{postdrift} into \eqref{secder} one obtains
\begin{eqnarray}
&&\bm{A}_{\rm s}\bm{\Sigma}^{\rm bb}\bm{A}_{\rm s} - \jm^{\rm bb\,\it{T}}\bm{A}_{\rm s} -\bm{A}_{\rm s}\jm^{\rm bb} -
\bm{A}_{\rm a}\big(\jm^{\rm bb}-\bm{\Sigma}^{\rm bb}\bm{A}_{\rm s}\big)\notag\\
&&- \big(\jm^{\rm bb\,\it{T}}-\bm{\Sigma}^{\rm bb}\bm{A}_{\rm s}\big)\bm{A}_{\rm a}
-\bm{W}=0.
\end{eqnarray}
The symmetric part of this determines
$\bm{A}_{\rm s}$, which is all we need for
\eqref{postdrift}, as
\be
\label{eq:riccati1app}
\bm{A}_{\rm s}\bm{\Sigma}^{\rm bb}\bm{A}_{\rm s} -\jm^{\rm bb\,\it{T}}\bm{A}_{\rm s} -\bm{A}_{\rm s}\jm^{\rm bb}=\bm{W}.
\ee
This is equation \eqref{eq:riccati1} in the main text -- we dropped the subscript ``s'' there -- and shows that 
the Lagrange multiplier $\bm{A}$ is identical to the (stationary) inverse backward covariance matrix, $\cbi$. 

\bibliography{Phdbib}

\end{document}